\documentclass[a4]{mn2e}\twocolumn
\usepackage{graphicx}
\usepackage[T1]{fontenc}

\begin{document}
\input{psfig}

\def\parn{\par\noindent}

\def\aeta{A\&A }
\def\aetal{A\&AL }
\def\aetas{A\&AS }
\def\apj{ApJ }
\def\apjs{ApJS }
\def\aj{AJ }
\def\mn{MNRAS }
\def\pasp{PASP }
\def\apjl{ApJL \rm}

\def\xp{\M{x}_\perp}
\def\los{line of sight}
\def\loss{lines of sight}
\def\ew{equivalent width}

\def\iint{\int\!\!\!\!\!\int}% double integral
\def\iiint{\iint\!\!\!\!\!\int}% triple integral
\def\ivnt{\iiint\!\!\!\!\!\int}% quadruple integral

\newcommand{\DXDYCZ}[3]{\left( \frac{ \partial #1 }{ \partial #2 }
                        \right)_{#3}}
\def\R#1{{\mathrm{#1}}}% roman font in math mode
\def\Sec#1{{Section~\ref{s:#1}}}
\def\Eq#1{{Eq.~\ref{e:#1}}}% equation reference
\def\eq#1{{eq.~\ref{e:#1}}}% equation reference
\def\Ep#1{{~(\ref{e:#1})}}% equation reference
\def\Eqs#1#2{{equations~(\ref{e:#1})-(\ref{e:#2})}}
\def\EQN#1{\label{e:#1}}        % eqn labelling a la Texsis
\def\Tab#1{{Table~\ref{t:#1}}}        % table reference
\def\Tap#1{{~\ref{t:#1}}}     % table no reference
\def\Fig#1{{Fig.~\ref{f:#1}}}% figure reference
\def\Figs#1#2{{Figs.~(\ref{f:#1})-(\ref{f:#2})}} % figure reference
\def\Fip#1{{~\ref{f:#1}}}% figure reference

\def\BLA#1{\noindent{\large\bf[$\bullet$ #1]}}
\def\Xtophe#1{\noindent{\bf[$\spadesuit$ #1]}}
\def\Bastien#1{\noindent{\bf[$\heartsuit$ #1]}}
\def\PPJ#1{\noindent{\bf[$\clubsuit$ #1]}}
\def\Steph#1{\noindent{\bf[$\diamondsuit$ #1]}}

\def\Lya{Ly-$\alpha$}

% equations reference
\def\M#1{{\mathbf{#1}}}% matrix notation
\def\T#1{{{#1}^{\bot}}}% transposition of a matrix
\def\d#1{{\R{d}{#1}}}% integrant
\def\mdot{\!\cdot\!}% matricial product
\def\MG#1{{\mbox{\boldmath $ #1$}}} %bold greek
\def\Mvariable#1{{\mathbf{#1}}}% matrix notation

\def\Mfunction#1{{\rm #1}}
\def\Mvariable#1{{\rm #1}}
\def\Muserfunction#1{{\rm #1}}

\def\tmop#1{{\rm #1}}
\def\etal{et al.\ }
\def\gtsima{$\; \buildrel > \over \sim \;$}
\def\ltsima{$\; \buildrel < \over \sim \;$}
\def\gsim{\lower.5ex\hbox{\gtsima}}
\def\lsim{\lower.5ex\hbox{\ltsima}}
\def\lesssim{\lsim}
\def\simgt{\lower.5ex\hbox{\gtsima}}
\def\simlt{\lower.5ex\hbox{\ltsima}}
\def\simpr{\lower.5ex\hbox{\prosima}}
\def\la{\lsim}
\def\ga{\gsim}
\def\CIV{C{\sc ~iv} }
\def\CIII{C{\sc ~iii} }
\def\HeIII{He{\sc ~iii} }
\def\HeII{He{\sc ~ii} }
\def\HI{H{\sc ~i} }
\def\SiIV{Si{\sc ~iv} }
\def\SiIII{Si{\sc ~iii} }
\def\MgII{Mg{\sc ~ii} }
\def\FeII{Fe{\sc ~ii} }
%these ones are for use in equations
\def\eCIV{{\rm C\, \sc iv}}
\def\eSiIV{{\rm Si\,\sc iv}}
\def\eMgII{{\rm Mg\, \sc ii}}
\def\eFeII{{\rm Fe\, \sc ii}}
\def\Lya{Ly$\alpha$~}
\def\zsun{{Z_\odot}}
\def\ie{{\frenchspacing\it i.e. }}
\def\eg{{e.g.\ }}
\newcommand{\be}{\begin{equation}}
\newcommand{\ba}{\begin{eqnarray}}
\newcommand{\ee}{\end{equation}}
\newcommand{\ea}{\end{eqnarray}}
\def\msun{\,{\rm M_\odot}}

%%%%%%%%%% Start TeXmacs macros
%%%%%%%%%% End TeXmacs macros

\title{The Sources of Intergalactic Metals}

\author[E.~Scannapieco \etal]{E.~Scannapieco$^1$, C.~Pichon$^{2,3}$,
B.~Aracil$^4$, P.~Petitjean$^{2,5}$, R. J. Thacker$^{6}$, \newauthor
D. Pogosyan$^7$, J. Bergeron$^2$, \& H. M. P. Couchman$^{8}$\\
%${}^1$ Osservatorio Astrofisico di Arcetri, 
%                     Largo E. Fermi 5, 50125 Firenze, Italy\\
${}^1$  Kavli Institute for Theoretical Physics,
 Kohn Hall, UC Santa Barbara, Santa Barbara, CA 93106 \\
${}^2$ Institut d'Astrophysique de Paris, 98 bis boulevard
       d'Arago, 75014 Paris, France \\
${}^3$ Observatoire de Strasbourg, 11 rue de
       l'Universit\' e, 67000 Strasbourg, France \\
${}^4$ Department of Astronomy, University of
       Massachusetts, Amherst, MA 01003\\
%${}^4$ Numerical Investigations in Cosmology (N.I.C.), CNRS, France.\\
${}^5$ LERMA, Observatoire de Paris, 61 avenue de l'Observatoire, F-75014
Paris, France\\
${}^6$ Department of Physics, Queen's University, Kingston, Ontario, K7L 3N6,
Canada\\
${}^7$ Department of Physics, University of Alberta,
  412 Avadh Bhatia Physics Laboratory, Edmonton, Alberta T6G 2J1, Canada\\
${}^8$ Department of Physics and Astronomy, McMaster University, 
1280 Main St.\ West, Hamilton, Ontario, L8S 4M1, Canada}
\date{Typeset \today ; Received / Accepted}
\maketitle

\begin{abstract}                    

We study the clustering properties of metals in the intergalactic
medium (IGM) as traced by 619 \CIV and 81 \SiIV absorption 
components with  $N \ge 10^{12}$ cm$^{-2}$  and  316 \MgII and 
82 \FeII absorption components with $N \ge 10^{11.5}$ cm$^{-2}$
in 19 high signal-to-noise (60-100 per pixel), high
resolution ($R=45000$) quasar spectra.   \CIV and \SiIV trace each
other closely  and their line-of-sight correlation functions $\xi(v)$
exhibit a steep decline at large separations and a flatter profile 
below $\approx 150$ km s$^{-1}$, with a large overall bias.
These features do not depend on absorber column densities,
although there are hints that the overall amplitude of $\xi_{\eCIV}$
increases with time over the redshifts range detected (1.5-3).
Carrying out a detailed smoothed particle
hydrodynamic simulation ($2 \times 320^3$, $57$ Mpc$^3$ comoving), we show that
the \CIV correlation function can not be reproduced by models in which
the IGM metallicity is a constant or a local function of overdensity
($Z \propto \Delta^{2/3}$).    However, the properties of  $\xi_{\eCIV}(v)$
are generally consistent with a model in which metals are confined within
bubbles with a typical radius $R_s$ about sources of mass $\geq M_s.$
We derive best-fit values of $R_s \approx 2$ comoving Mpc and $M_s
\approx 10^{12} \msun$ at $z=3$.  Our lower redshift ($0.5-2$)
measurements of the \MgII and \FeII correlation functions also
uncover a steep decline at large separations and a flatter profile at
small  separations, but the clustering is even higher than in the
$z=$1.5-3 measurements, and the turn-over is shifted to somewhat smaller
distances $\approx 75$ km s$^{-1}$.  
Again these features do not change with column density,
but there are hints that the amplitudes of $\xi_{\eMgII}(v)$ and
$\xi_{\eFeII}(v)$  increase with time. 
We describe an analytic ``bubble'' model for these species,
which come from regions that are too compact to be accurately
simulated numerically, deriving best-fit values of 
$R_s \approx 2.4$  Mpc and $M_s \approx 10^{12} \msun.$
Equally good analytic fits to all four species are found in a 
similarly biased high-redshift enrichment model 
in which metals are placed within $2.4$ comoving Mpc of
$M_s \approx 3 \times 10^9$ sources at $z = 7.5.$

\end{abstract}     

%\begin{keywords}{cosmology: observations - 
%    intergalactic medium - quasars: absorption lines - 
%large-scale structure of the universe}               
%\end{keywords}

\section{Introduction}

Pollution is ubiquitous.  Even in  the
tenuous intergalactic medium (IGM), quasar (QSO) absorption line
studies have encountered heavy elements in all regions in which they
were detectable (Tytler \etal 1995; Songaila \& Cowie 1996).  Such analyses
were limited at first to somewhat overdense regions of space, traced
by \Lya clouds with column densities $N_{\rm HI} \geq 10^{14.5}$ cm$^{-2}$.  
Here measurements of $N_{\rm CIV}/N_{\rm HI}$ indicated
that typically $[C/H] \simeq -2.5$ at $z \simeq 3$, with an order of
magnitude scatter (Hellsten \etal 1997; Rauch, Haehnelt, \& Steinmetz
1997).

Pushing into more tenuous regions, 
statistical methods have shown that unrecognized weak
absorbers must be present in order to reproduce the global \CIV
optical depth (Ellison \etal 2000), and that a minimum IGM metallicity
of approximately $3\times 10^{-3} Z_\odot$ was already in place at $z
= 5$ (Songaila 2001; hereafter S01).  While the filling factor
of metals in such tenuous structures is an object of intense
investigation and debate (Schaye \etal 2000; Petitjean 2001; Bergeron
\etal 2002; Carswell \etal 2002; Simcoe \etal 2002; Pettini \etal
2003; Schaye \etal 2003; Aracil \etal 2004) their very existence has 
profound cosmological implications.

As the presence of metals increases the number of lines available for
radiative cooling, even modest levels of enrichment can
greatly enhance the cooling rate (e.g.\ Sutherland \& Dopita 1993),
which
has the potential to accelerate the formation of massive ($\gsim
10^{12} \msun$) galaxies  (e.g.\ Thacker, Scannapieco \& Davis 2002).
Furthermore, significant  preenrichment is necessary to
reproduce the abundances of G-dwarf stars in the
Milky Way (e.g.\ van de Bergh 1962; Schmidt 1963) and nearby galaxies
(e.g.\ Thomas, Greggio, \& Bender 1999).

Similarly, the violent events that propelled heavy elements 
into the space between galaxies have important
implications for the thermal and velocity structure of the IGM (e.g.\
Tegmark, Silk, \& Evrard 1993; Gnedin \& Ostriker 1997; Cen \& Bryan
2001).  Outflows energetic enough to eject metals from the
potential wells of dwarf galaxies, for example, would have exerted
strong feedback effects on nearby objects (Thacker, Scannapieco, \&
Davis 2002).  In this case the winds impinging on pre-virialized
overdense regions would have been sufficiently powerful to strip the
baryons from their associated dark matter, greatly
reducing the number of $\lsim 10^{10} \msun$ galaxies formed
(Scannapieco, Ferrara, \& Broadhurst 2000; 
Sigward, Ferrara, \& Scannapieco 2005).
 
Yet despite their many consequences, the details of how metals came 
to enrich the IGM are unclear.
While numerous starburst-driven outflows have
been observed at $z = 3$ (Pettini \etal 2001) and in lensed galaxies
at $4 \lesssim z \lesssim 5$ (Frye, Broadhurst, \& Benitez 2002), it
is unclear whether these objects are responsible for the majority of
cosmological enrichment.  In fact a variety of theoretical arguments
suggest that such galaxies represent only the tail end of a larger
population of smaller ``pre-galactic'' starbursts that mostly formed
at much higher redshifts (Madau, Ferrara, \& Rees 2001; Scannapieco,
Ferrara, \& Madau 2002).  
On the other hand, active galactic nuclei
are observed to host massive outflows (Begelman, Blandford,
\& Rees 1984; Weyman 1997), whose contribution from less luminous
objects at intermediate redshifts remains unknown (e.g. Fan \etal
2001).  The impact of such lower-redshift events on the IGM is also
hinted at by the ``stirring'' of \CIV systems observed in studies of
lensed QSO pairs (Rauch, Sargent, \& Barlow 2001).  Finally, a number
of theoretical studies suggest that primordial,
metal-free stars may have been very massive (e.g.\ Bromm \etal 2001;
Schneider \etal 2002), resulting in a large number of tremendously
powerful pair-production supernovae, which  distributed metals into the
IGM at extremely early redshifts $\gsim 15$ 
(Bromm 2003; Norman, O'Shea, \& Paschos 2004).

While perhaps the main feature shared by such scenarios is their
dependence on a poorly-understood population of presently undetectable
objects, this assessment paints an overly bleak picture. 
Regardless of which 
objects enriched the IGM, it is clear that they must have formed in the
densest regions of space, regions that are far more clustered than the
overall dark matter distribution.  
Furthermore this ``geometrical biasing''
is a systematic function of the masses of these structures, an effect
that has been well studied analytically and numerically (e.g.\ Kaiser
1984; Jing 1999).  Thus the observed large-scale clustering of
metal absorbers encodes valuable information about the masses of the
objects from which they were ejected.
Likewise, as the maximal extent of each
enriched region is directly dependent on the velocity at which the
metals were dispersed, measurements of the small-scale clustering of
these absorbers are likely to constrain the energetics of their sources.

Previous studies of the two-point correlation function of \CIV 
components have shown that they cluster strongly on velocity scales  up
to 500~km~s$^{-1}$ (Sargent et al. 1980, Steidel 1990, Petitjean \&
Bergeron 1994, Rauch et al. 1996).  It has often been suggested that
the clustering signal reflects a combination of (i) relative motions
of clouds within a galactic halo and (ii) galaxy clustering.  More
recently Boksenberg, Sargent, \& Rauch (2003; hereafter BSR03) 
have gathered a sample of 908 \CIV
absorber components clumped into 199 systems in the redshift range
1.6~$<$~$z$~$<$~4.4 identified in the Keck spectra of nine QSOs. 
They conclude that most of the signal is due to the clustering of
components within each system, where a system is defined as a set of
components that is ``well-separated'' from its neighbours
as identified by the observer.  In this case almost all the systems 
extend less than 300~km~s$^{-1}$ and most extend less than 150~km~s$^{-1}$.
They did not observe clustering between systems on the larger scales 
expected for galaxy clustering, although they concluded from their 
measurements of component clustering and ionisation balance 
that each system was closely associated with a galaxy. 

In Pichon \etal (2003; hereafter Paper I) we used 643 \CIV and 104
\SiIV absorber components, measured by an automated procedure in 19
high signal-to-noise quasar spectra, to place strong constraints on
the number and spatial distribution of intergalactic metals at
intermediate redshifts ($2 \leq z \leq 3$).  In this work, we 
showed that the correlation functions of intergalactic \CIV and \SiIV 
could be understood in terms of the clustering of metal bubbles of a typical
comoving radius $R_s$ around sources whose biased clustering was
parameterized by a mass $M_s$.  A similar picture was also put 
forward in BSR03, but in our case significant large-scale 
clustering, similar to that seen in galaxies, was observed.

In this paper we extend the analysis in Paper I in three important ways.  First
we carry out a more detailed study of the physical properties of \CIV
and \SiIV absorbers and the relationship between local quantities and
the overall spatial distribution.  Second, we carry out a similar
analysis of \MgII and \FeII absorbers in our observational
sample, which probe the IGM in a somewhat lower redshift range.
Finally, we replace our dark-matter only modeling of Paper I with a
full-scale smoothed-particle hydrodynamical simulation.  We then
generate simulated metal-line spectra by painting bubbles of
metals directly onto the gas distribution at $z \geq 2$.  By analyzing
the resulting spectra with the same automated procedure applied to the
measured data-set, we are able to place our models and observations on the
same footing, drawing important constraints on the sources of metals.
Motivated by measurements of the cosmic microwave background, the number 
abundance of galaxy clusters, and high-redshift supernovae (e.g.\
Spergel et al.\ 2003; Eke \etal 1996; Perlmutter \etal 1999)  we adopt 
cosmological parameters of $h=0.7$, $\Omega_m$ = 0.3, 
$\Omega_\Lambda$ = 0.7, $\Omega_b = 0.044$, throughout this investigation
where $h$ is the Hubble Constant in units of 100 km s$^{-1}$ Mpc$^{-1}$
and $\Omega_m$, $\Omega_\Lambda$, and
$\Omega_b$ are the total matter, vacuum, and baryonic densities in units
of the critical density, $\rho_{\rm crit}.$

The structure of this work is as follows.  In \S2 we summarize the
properties of our data set and reduction methods. In \S3 we present
the number densities of {C{\sc ~iv}}, {Si{\sc ~iv}}, {Mg{\sc ~ii}}, and
{Fe{\sc ~ii}}, and estimate the cosmological densities of these species.  
In \S4 we study the spatial clustering of these species and how
it is related to local quantities such as column density and
abundance ratios.  In \S5 we describe our numerical model for the
distribution of neutral hydrogen in the IGM and compare it
with observations. In \S6 we extend our method to include various
models of cosmological enrichment and in \S7 we compare these models
to the observed distribution of \CIV to derive constraints
on the sizes and properties of sources of cosmological metals.  In \S8
we discuss an analytic model that is particularly suitable for
comparisons with  the distribution of \MgII and  {Fe{\sc ~ii}}, as
numerical analyses of these species are beyond the capabilities of our 
simulation. Conclusions are given in \S9.

\section{Data Set and Analysis Methods}
 
\subsection{Data and Reduction}

The ESO Large Programme ``The Cosmic Evolution of the IGM'' was
devised to provide a homogeneous sample of QSO sight-lines suitable
for studying the Lyman-$\alpha$ forest in the redshift range
1.7$-$4.5.  High resolution ($R$~$\approx$~45000), high signal-to-noise
(60-100 per pixel) spectra were taken over the wavelength ranges
3100--5400 and 5450--10000~\AA, using the UVES spectrograph on the
Very Large Telescope (VLT). Emphasis was given to lower redshifts to
take advantage of the very good sensitivity of UVES in the blue and
the fact that the Lyman-$\alpha$ forest is less blended.  The
distribution of redshifts, and the resulting coverage of various metal
line absorbers are given in Table 1.  In all cases we consider only
metal absorption lines redward of the \Lya forest, to avoid the
extensive blending in this region, and blueward of 8110 \AA,  to avoid
contamination from sky lines.  The regions between  5750-5830 \AA,
6275-6323 \AA,  6864-6968 \AA,  7165-7324 \AA, and 7591-7721 \AA \, were
also excluded from our sample due to sky-line contamination.  The
{C{\sc ~iv}}, {Si{\sc ~iv}}, {Mg{\sc ~ii}}, \& \FeII metal lines
discussed in this paper were  well-detected over the redshift  ranges
of 1.5-3.0, 1.8-3.0, 0.4-1.8, and 0.5-2.4 respectively.

\begin{table*}
\centerline{
\begin{tabular}{llccccc}
\hline \qquad Name & \qquad & \multicolumn{5}{c}{Coverage} \\   &
$z_{\rm em}$           &   Forest    & C~{\sc iv} & Si~{\sc iv}
&	 Mg~{\sc ii} & Fe~{\sc ii}  \\ \hline\\ PKS~2126$-$158  &
3.280 & 2.61$-$3.28  & 2.36$-$3.28  & 2.74$-$3.28	&
0.85$-$1.85	& 1.03$-$2.42\\ Q~0420$-$388    & 3.117 & 2.47$-$3.12
& 2.23$-$3.12  & 2.59$-$3.12 	& 0.79$-$1.85 	& 0.95$-$2.42 \\
HE~0940$-$1050  & 3.084 & 2.45$-$3.08  & 2.21$-$3.08  & 2.56$-$3.08
& 0.77$-$1.85  	& 0.93$-$2.42\\ HE~2347$-$4342  & 2.871 & 2.27$-$2.87
& 2.04$-$2.87  & 2.38$-$2.87	& 0.68$-$1.85	& 0.83$-$2.42\\
HE~0151$-$4326  & 2.789 & 2.20$-$2.79  & 1.97$-$2.79  &
2.31$-$2.79	& 0.64$-$1.85	& 0.79$-$2.42\\ Q~0002$-$422    &
2.767 & 2.18$-$2.77  & 1.96$-$2.77  & 2.29$-$2.77	&
0.64$-$1.85	& 0.78$-$2.42\\ PKS~0329$-$255  & 2.703 & 2.13$-$2.70
& 1.91$-$2.70  & 2.23$-$2.70	& 0.61$-$1.85	& 0.75$-$2.42\\
Q~0453$-$423    & 2.658 & 2.09$-$2.66  & 1.87$-$2.66  &
2.19$-$2.66	& 0.59$-$1.85	& 0.73$-$2.42\\ HE~1347$-$2457  &
2.611 & 2.05$-$2.61  & 1.83$-$2.61  & 2.15$-$2.61	&
0.57$-$1.85	& 0.70$-$2.42\\ HE~1158$-$1843  & 2.449 & 1.91$-$2.45
& 1.71$-$2.45  & 2.01$-$2.45	& 0.50$-$1.85	& 0.63$-$2.42\\
Q~0329$-$385    & 2.435 & 1.90$-$2.44  & 1.70$-$2.44  & 2.00$-$2.44
& 0.49$-$1.85 	& 0.62$-$2.42\\ HE~2217$-$2818  & 2.414 & 1.88$-$2.41
& 1.68$-$2.41  & 1.98$-$2.41 	& 0.48$-$1.85 	& 0.61$-$2.41\\
Q~1122$-$1328 & 2.410 & 1.87$-$2.41 & 1.68$-$2.41  & 1.98$-$2.41   &
0.39$-$1.85  	& 0.61$-$2.41\\       Q~0109$-$3518   & 2.404 &
1.87$-$2.40  & 1.67$-$2.40  & 1.97$-$2.40 	& 0.48$-$1.85 	&
0.61$-$2.40\\ HE~0001$-$2340  & 2.263 & 1.75$-$2.26  & 1.56$-$2.26  &
1.84$-$2.26 	& 0.42$-$1.85 	& 0.54$-$2.26 \\ PKS~0237$-$23   &
2.222 & 1.72$-$2.22  & 1.53$-$2.22  & 1.81$-$2.22 	& 0.40$-$1.85
& 0.53$-$2.22 \\ PKS~1448$-$232  & 2.220 & 1.72$-$2.22  & 1.53$-$2.22
& 1.81$-$2.22	& 0.40$-$1.85	& 0.52$-$2.22\\ Q~0122$-$380    &
2.190 & 1.70$-$2.19  & 1.50$-$2.19  & 1.78$-$2.19	&
0.38$-$1.85	& 0.51$-$2.19\\ HE~1341$-$1020  & 2.135 & 1.65$-$2.14
& 1.46$-$2.14  & 1.74$-$2.14	& 0.36$-$1.85	& 0.48$-$2.14\\
\hline\\
\end{tabular}}
\label{objects}
\caption {List of lines of sight.  Here $z_{{\rm em}}$ is the quasar
redshift, while \Lya forest is used only redward of the Ly$\beta$
transition at $1025.7$ \AA, and metal absorption lines are used only
redward of the {\Lya} forest and blueward of 8130 \AA.}
\end{table*}

Observations were performed in service mode over a period of two
years. The data were reduced using the UVES context of the ESO MIDAS
data reduction package, applying the optimal extraction method, and
following the pipeline reduction step by step. The extraction slit
length was adjusted to optimize sky-background subtraction. While
this procedure systematically underestimates the sky-background
signal, the final accuracy is better than 1\%.  Wavelengths were
corrected  to vacuum-heliocentric values and individual 1D spectra
were combined  using a sliding window and weighting the signal by the
total errors in each pixel.

The underlying emission spectrum of each quasar was estimated using an
automated iterative procedure that minimizes the sum of a
regularisation term and a $\chi^2$ term that was computed from the
difference between the quasar spectrum and the continuum estimated
during the previous iteration.  Finally the spectrum was divided by
this continuum, leaving only the information relative to absorption
features.

\subsection{Metal Line Identification}

Metal-line absorbers  were   identified  using  an  automated  two-step
procedure.   For  each species  that  has  multiple transitions,   we
estimated  the  minimal flux compatible with the data for all pixels
of the  spectrum.  This was done by first finding the pixels associated
with the transition wavelengths $w_i$  of a given species and then
taking  the maximum of the flux  values in   these pixels, scaled by
$w_i f_i$, where $f_i$ is the oscillator strength
associated with each of the transitions.

A standard detection threshold was then applied to these spectra, such
that only absorption features with equivalent widths (EWs) larger than
5 times the noise rms were  accepted, giving a first list of possible
identifications.   This list was cleaned, using the
similarity of the profiles of the transitions of a species and 
applying  simple physical criteria that correlate  the detection of
two  different species. For  instance, one  criterion implies
that the  detection of a \SiIV  system at a given  redshift should be
associated with the detection of a \CIV system.

Next, each system was fitted with Voigt profiles,  taking care of their
identification and possible blends with other systems. The first guess
and the  final Voigt  profile decomposition were  carried out  using the
VPFIT software  (Carswell \etal 1987). Our decomposition 
of saturated systems is conservative, in that
it introduces additional unsaturated  components only if there is some 
structure in the 1551 $\AA$ line that reveals their presence.
This fitting procedure is described in
detail in Aracil  \etal  (2005)  and has  been  tested on  simulated
spectra, doing well  for all components with realistic  values of $N$ and
$b$.

Finally we applied a set of five cuts to the automated list generated
by VPFIT: $\log N \, [{\rm cm}^{-2}] \, \geq 12$ for \CIV and \SiIV
and $\log N \, [{\rm cm}^{-2}] \, \geq 11.5$ for \MgII and \FeII
due to the detection limit
of our procedure, $b \geq 3 $ km s$^{-1}$ to avoid false detections due to
noise spikes, $\log N [{\rm cm}^{-2}] \leq 16$ to remove very badly
saturated components, and $b \leq 45$ km s$^{-1}$ to avoid false detections
due to errors in continuum fitting.  For the analyses presented
here, we removed all associated components within $5000$ km s$^{-1}$ of the 
quasar redshifts. These cuts resulted in a final data
set of 619 \CIV (1548 \AA, 1551 \AA), 
       81 \SiIV (1394 \AA, 1403 \AA), 
       316 \MgII (2796 \AA, 2803 \AA),
       and 82 \FeII (2344 \AA, 2473 \AA, 2382 \AA) components,
drawn from 688, 102, 320, and 88
components
respectively, if we include the associates. These numbers differ 
slightly from those presented in Paper I due to further 
refinements in our detection procedure.

\section{Number Densities}

We first used our sample to compute  the  column density  distribution
function, $f(N),$ again working in the above assumed cosmology.
Following  Tytler (1987), $f(N)$ is defined as the number
of  absorbing components per  unit column  density and  per unit  redshift 
path, ${\rm d}X.$  In this paper, we adopt a definition of 
${\rm d}X \equiv (1+z)^2  \left[ \Omega_\Lambda+ \Omega_m (
1+z)^3 \right]^{-1/2}  dz$ such that at all redshifts $f(N)$ 
does not evolve for a population whose physical size and
comoving space density are constant.  
Note that this definition is slightly different from that used in Paper I and
in S01, namely ${\rm d}X'  \equiv (1+z)^{1/2} dz$, although 
when $z > (\Omega_\Lambda/\Omega_m)^{1/3} - 1 = 0.32,$  as is 
appropriate for our sample, ${\rm d}X'$ can be very closely approximated as
$\Omega_m^{1/2} {\rm d}X$ for comparison with previous analyses.

\begin{figure}
\centerline{\psfig{figure=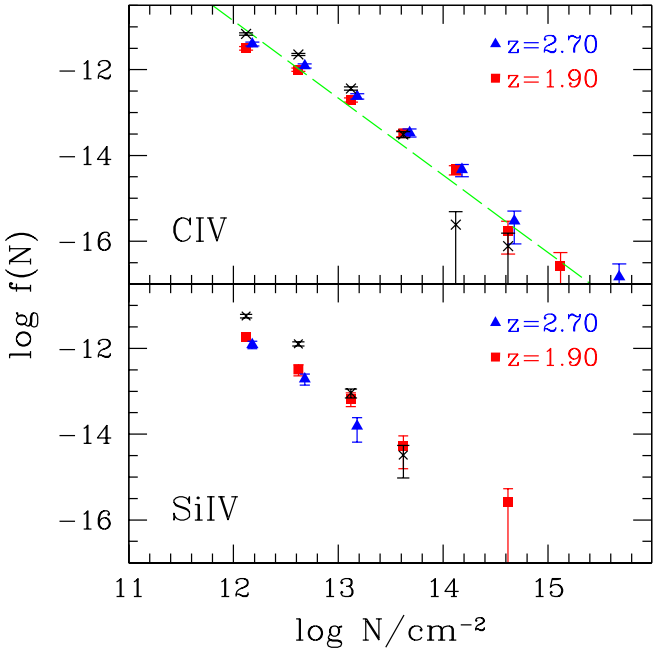,height=9.5cm}}
\caption{Column density distribution of \CIV (upper panel) and \SiIV
(lower panel) absorption components.  In each panel, components are
divided into two redshift bins: $1.5 \leq z \leq 2.3$ (squares) and
$2.3 \leq z \leq 3.1$ (triangles).  Column density bins are $10^{0.5}
N$ cm$^{-2}$ wide and error bars in this and all further plots are 
1-$\sigma$.   The dashed line is the power law fit measured in S01.  
Finally the small crosses 
are the full set of \CIV and \SiIV components 
identified by  BSR03, with our cuts imposed.}
\label{fig:abundance}
\end{figure}

In Figure \ref{fig:abundance} we plot $f(N)$ for both \CIV and \SiIV
components, as was presented in Paper I.   The mean redshifts of \CIV
and \SiIV in our sample were 2.16 and  2.38, respectively, and so in
this plot we  divide the data into two redshift bins from $1.5 \leq z
\leq 2.3$ and $2.3 \leq z \leq 3.1.$  Both species are consistent with
a lack of redshift evolution, as found by previous lower-resolution
studies of \CIV and \SiIV (S01; Pettini \etal 2003), and
pixel-by-pixel analyses of intergalactic \CIV (Schaye \etal 2003).
The overall density distribution of \CIV is also consistent with a
power-law of the form $f(N) = B N^{-\alpha}$ with $\alpha=1.8$ and
$\log_{10} f = -12.7$ at $10^{13}$ cm$^{-2}$ as fitted  by S01.  Finally,
we compare our results with the dataset collected in BSR03
from nine QSO spectra with a S/N $\approx 50$ per pixel. 
Here and below we use the {\em full data set} taken by BSR03, to which
we apply
exactly the same cuts as we do to our data.  
For components with columns $\approx 10^{13}$ cm$^{-2}$ these data
sets are quite similar.  However, a significant
difference between this sample and our own is the fit to the 
saturated \CIV components with $\log(N_\eCIV) \geq 14$.  These have
been decomposed into a large number of smaller  $\log(N_\eCIV) \leq 12.5$
systems in the BSR03 analysis, while our decomposition only
introduces additional unsaturated components if there is 
structure in the 1551 $\AA$ line. Extrapolating the results of Songaila 
(2001) to column depths below $10^{13}$ cm$^{-2}$ also yields a
distribution similar to ours.

While fewer in total, the \SiIV components in the lower panel are also
consistent with a lack of evolution, following a similar power law
with a lower overall magnitude.   Note that in this figure the error
bars are purely statistical, estimated as one over the square-root of
the number of components in each bin.  Again, for comparison, we include 
the number densities computed from the full BSR03 sample, with our 
cuts applied.  While this comparison is noisier,
the overall trends are the same: at $10^{13}$ cm$^{-2}$ the number
densities are similar, while saturated components are decomposed
into a larger number of smaller systems in the BSR03 data set.

\begin{figure}
\centerline{\psfig{figure=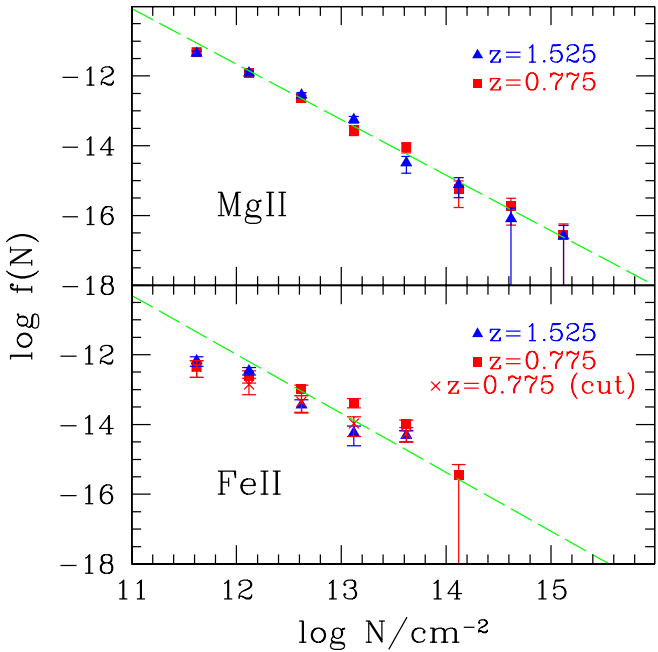,height=9.5cm}}
\caption{Column density distributions of \MgII (upper panel) and \FeII
(lower panel) absorption components.  In each panel, components are
divided into two redshift bins: $0.4 \leq z \leq 1.15$ (squares) and
$1.15 \leq z \leq 1.9$ (triangles).  As in Fig.\ 1, the column density 
bins are $10^{0.5} N$ cm$^{-2}$ wide.   The dashed lines 
correspond to the power-law fits described in the text, and in the
lower panel we also include $f$ values when the large 
$z = 0.836$ system in Q0002-422 is removed (see text).}
\label{fig:abundance2}
\end{figure}

In Figure \ref{fig:abundance2} we plot $f(N)$ for both \MgII and
{Fe{\sc ~ii}}, now going down to a minimum column density of $10^{11.5}$
cm$^{-2}$ which corresponds to roughly the same optical depth as
$10^{12}$ cm$^{-2}$ for \CIV and {Si{\sc ~iv}}.  For \MgII and \FeII
the relevant doublets  are at substantially longer restframe
wavelengths, and therefore our UVES  detections primarily occur at
lower redshifts.  Thus the mean redshifts of \MgII and \FeII are only
1.05, and 1.38, and we divide our data into  bins from $0.4 \leq z
\leq 1.15$ and $1.15 \leq z \leq 1.9.$   These lines arise in lower
ionization gas and are often thought of as tracers of quiescent
clouds,  probably associated with galaxies (\eg Petitjean \& Bergeron
1990; Churchill \etal 1999;  Churchill, Vogt, \& Charlton 2003).

Like its higher-ionization counterparts, \MgII is  consistent with a
lack of evolution in number densities over the observed redshift
range.  In the \FeII case, however, a significant excess of
intermediate  column density components is found at lower redshifts.  A
closer inspection of the data indicates that this feature is caused by
a single large system in Q0002-422, at $z = 0.836$, which spans over
560 km s$^{-1}$.    The removal of this system results in the third
set of points in the lower panel of Fig.\ \ref{fig:abundance2}, which
are consistent with the higher-redshift values.  The large impact of
this system in our measurements suggests that simple $\sqrt{N}$
estimates may somewhat  underpredict the statistical error on our
measurement.  This hints at strong clustering between \FeII
components, which is in fact  measured, as we discuss in detail below.

Statistical fluctuations aside,  the overall density distributions of
\MgII and \FeII are largely consistent with the power law fits
obtained from previous measurements, apart from showing only  a weak
deviation in the lowest $N_{\eFeII}$ bin, probably due to incompleteness.   
In this case, the dashed-lined fits in Fig \ref{fig:abundance2} are
$f(N) = B N^{-\alpha}$ with $\alpha=1.6$ and $\log_{10} f = -13.2$ at
$10^{13}$ cm$^{-2}$ for \MgII and  $\alpha=1.7$ and $\log_{10} f =
-13.4$ at $10^{13}$ cm$^{-2}$  for {Fe{\sc ~ii}}.
While some flattening of $f_{\eMgII}(N)$ at even higher columns
is necessary to match observations at column densities $\ge 10^{16.5}$ cm$^{-2}$
(Prochter, Prochaska, \& Burles 2004), for the column densities
in our sample our measured slopes are identical
with those determined by previous studies.  In particular
our $\alpha$ fits match those of
Churchill, Vogt, \& Charlton (2003),
although our $B$ values are different as these authors did not
attempt to normalize their results by the total redshift path 
observed.

In summary, our automatic identification procedure produces a set of components
whose column density distributions are consistent with previous measurements,
 complete to  $N \gsim 10^{12}$ cm$^{-2}$ for \CIV and {Si{\sc ~iv}}, 
and complete to $N \gsim 10^{11.5}$ cm$^{-2}$ for \MgII and {Fe{\sc ~ii}}.
No evolution in $f$ is seen for any species over the full redshift
range probed, indicating that the majority of IGM enrichment is likely to 
have occurred before the redshifts observed in our sample.

Finally, our number densities allow us to compute the total 
cosmological densities of each of the detected species.
Following S01, we express these in terms of a mass fraction
relative to the critical density, which can be computed 
as
\be
\Omega_{\rm ion} = \frac{H_0 m_{\rm ion}}{c \rho_{\rm crit}}
\frac{\sum N_{\rm ion}}{\Delta X} = 1.4 \times 10^{-23} A 
\frac{\sum N_{\rm ion}}{\Delta X},
\ee
where $H_0$ is the Hubble constant, $m_{\rm ion}$ is the mass
of the given ion, $A$ is its atomic number, and $\Delta X$ is
the total redshift path over which it is measured.   
The results of this analysis are given in Table 2.  Note that
these values are {\em species} densities, and no ionization
corrections have been applied to estimate the corresponding
element densities.   Again these values are broadly consistent
with previous measurements, although there is a significant scatter
due to the fact that most of the material lies in the largest, rarest
components.  Thus previous studies have found 
$\Omega_{\eCIV}$ values as disparate as
$6.8 \times 10^{-8}$ at $z=2.5$ (S01),  $3.8 \pm 0.7 \times 10^{-8}$ 
(BRS03), and between $3.5 \times 10^{-8}$ and $7.9 \times 10^{-8}$
depending on the method of analysis (Simcoe, Sargent, \& Rauch 2004).

\begin{table*}
\centerline{
\begin{tabular}{lllll}
\hline
Species & $\left< z \right>$  & log(N/cm$^{-2}$)& $\Omega$ & $\Delta \Omega$ \\
\hline\\
\CIV   & 2.2  & 12-16 & $7.54 \times 10^{-8}$ & $\pm  2.16  \times 10^{-8}$\\
\SiIV  & 2.4  & 12-16 & $6.00 \times 10^{-9}$ & $\pm  1.21  \times 10^{-9}$\\
\MgII  & 1.1  & 11.5-16 & $5.95 \times 10^{-8}$ & $\pm  2.23  \times 10^{-8}$\\
\FeII  & 1.4  & 11.5-16 & $1.87 \times 10^{-8}$ & $\pm  0.36  \times 10^{-8}$\\
\hline\\
\end{tabular}}
\label{tab:metaldensity}
\caption{Cosmological densities of detected species.}
\end{table*}

\section{Spatial Distribution}

\subsection{Carbon IV and Silicon IV}

Having constructed a sample of well-identified metal absorption
components, we then computed their two-point correlation function in
redshift space, $\xi(v)$.  This quantity was previously studied in
Rauch \etal (1996) who noted a marked similarity between $\xi(v)$ of
\CIV and {Mg{\sc ~ii}}, in BSR03, who carried out a two-Gaussian fit
(see also Petitjean \& Bergeron 1990, 1994), and in Paper I.
For each
quasar, we computed a histogram of all velocity separations and
divided by the number expected for a random distribution.  
Formally, the correlation function for a QSO $\ell$ is 
\be
\xi^{\ell}(v_k) + 1 = \frac{n^{\ell}_k}{\left< n^{\ell}_k \right>},
\ee
where $n^\ell_k$ is the number of pairs separated by a velocity
difference corresponding to a bin $k$, and $\left< n^\ell_k \right>$ is
the average number of such pairs that would be found in the redshift
interval covered by QSO $\ell$, given a random distribution of
redshifts with an overall density equal to the mean  density in the
{\em sample}.  Alternatively, we may consider all QSOs  at once and compute
\be
\xi(v_k) + 1 = 
\frac{\sum_\ell n^{\ell}_k}{\sum_\ell \left< n^{\ell}_k \right>},
\ee
or equivalently
\be
\xi(v_k) + 1 
=  \sum_{\ell} w^{\ell}_k \, [\xi^\ell(v_k)+1] \quad \tmop{with} \quad
   w^{\ell}_k \equiv 
   \frac{\left< n^{\ell}_k \right>}{\sum_\ell \left< n^{\ell}_k \right>},
\ee
that is weighting the correlation found for each QSO by the number of 
random pairs that are expected given the redshift coverage of that QSO.
The statistical variance in this measurement is given by
\be
\sigma^{2}_k  =  \sum_{\ell} (w^{\ell}_k)^2 \sigma^{2,\ell}_k,
\ee
where $\sigma^{2,\ell}_k$ is the variance associated with 
bin $k$ of quasar $\ell.$
In Paper I, we estimated this quantity according to the usual formula
\be
\sigma^{2,\ell}_k =
\frac{n^{\ell}_k}{\left< n^{\ell}_k \right>^2},
\ee
which gives the Poisson error in our measurement.  In the
results presented here, however, we adopt a more conservative
approach, and  also include the additional 
scatter caused by the finite sample size used to construct
the correlation function (Mo, Jing, \& B\" orner 1992). In this case 
\be
\sigma^{2,\ell}_k = 
\frac{1} {\left< n^{\ell}_k \right>^2}
\left[n^{\ell}_k + 4\frac{(n^{\ell}_k)^2}{N^\ell}  \right],
\ee
where $N^{\ell}$ is the total number of components detected in QSO $\ell$.
Note that the presence of this additional scatter highlights the strength of
our high signal-to-noise data set, as it allows us to work in the limit
in which  the number of \CIV components detected in {\em each quasar} is 
large.

\begin{figure}
\centerline{\psfig{figure=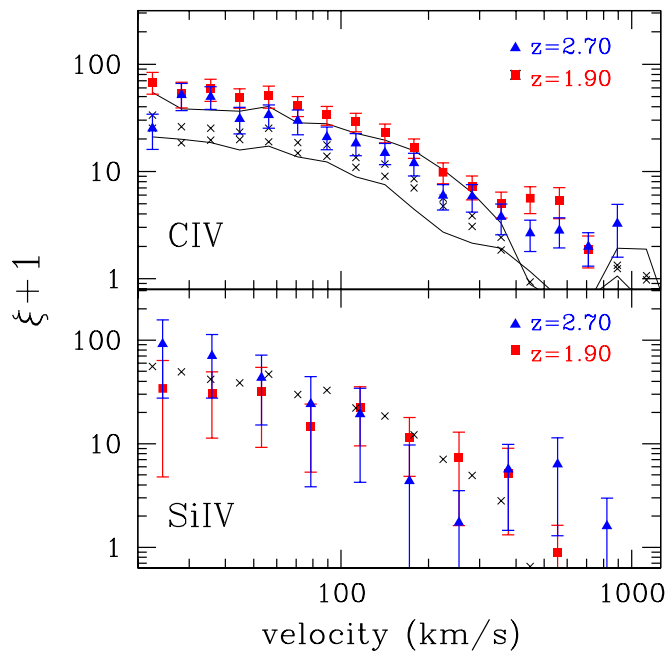,height=9.5cm}}
\caption{Two point correlation function of \CIV (upper panel) and
\SiIV (lower panel) absorption components.  In each panel the components
have been divided into two redshifts bins, with symbols as in Fig.\ 1.
The upper panel also includes a number of comparisons with previous
measurements.  In particular: the lower set of crosses corresponds to
the full set of components defined in BSR03, normalizing  each 
QSO individually;
the upper set of crosses corresponds to imposing a column density cut of
$10^{12}$ cm$^{-2}$, normalizing each QSO by its expected
number of pairs;
and the solid lines correspond to dividing the 
BSR03 data into a subsamples with $z \geq 3.1$ (lower line) and $z < 3.1$
(upper line), as described in the text. 
In the lower panel, the crosses corresponds to imposing a column density cut 
of $10^{12}$ cm$^{-2}$ on the \SiIV components observed in BSR03 
and normalizing each QSO by its expected number of pairs.}
\label{fig:CIV}
\end{figure}

The resulting correlation functions are shown in Fig.\ \ref{fig:CIV},
again split into two redshift bins. Interestingly, in the better-measured
\CIV case, there are hints that the $z \leq 2.3$ correlation function may
be enhanced with respect to the high-redshift one.
Furthermore, this growth is consistent with a population of absorbers
that ``passively'' evolves by following the motion of the IGM
during the formation of structure, as we discuss in further detail
in \S8.

In the upper panel of this figure we also plot  correlation functions
computed from the sample defined in BSR03, which is drawn  from the
spectra of nine QSOs with a mean redshift of 3.1  and a
signal-to-noise per pixel $\approx 50.$ In this case we show results
obtained both from using the full data set, normalizing  each quasar
individually  (as was carried out in BSR03),  and from imposing a
lower cut-off at $N_{\eCIV,\rm min} = 10^{12}$ cm$^{-2},$ normalizing
each quasar by the {\em expected} number of pairs (as was carried out
in our analysis).  In both cases the resulting $\xi_{\eCIV}$ values
are similar and somewhat lower in amplitude than our measurements.
Rauch \etal (1996) similarly have found a lower amplitude.  Dividing
the BSR03 data into a $z < 3.1$ bin with a mean redshift of 2.5 and a
$z > 3.1$ bin with a mean redshift of 3.6 resulted in correlation
functions given by  the solid curves (again calculated according to
our method).  Furthermore the amplitude of the $z=2.5$ BSR03
correlation function is similar to our measurements, which are drawn
from a sample with a mean redshift of $2.3.$   However, the
higher-redshift curve is substantially lower, again indicating that
$\xi_{\eCIV}$ is likely to evolve with redshift.  This was also 
suggested by the analysis in BSR03 Figure 14, although they
point out the changing ionizing background may also be an issue.
Finally we note that while the BSR03
sample shows a relative lack of components at   $\approx 500$ km/s.
This is very near the \CIV  doublet separation.

Moving to the bottom panel, we see that  the overall shape and
amplitude of the \CIV and \SiIV correlation functions are similar and
are consistent to within the \SiIV measurement errors, as was
discussed in Paper I.  Both functions exhibit a steep decline at large
separations and a flatter profile at small separations, with an elbow
occurring at $\approx 150$ km s$^{-1}$.  Both functions are also
consistent with the correlation one obtains from the full
BSR03 \SiIV sample, after applying our cuts.  Finally, as was noted in
Paper I, there is a weak low redshift feature at $\approx 500$ km
s$^{-1}$ in $\xi_{\eCIV}$, the origin of which we explore in \S 4.2.

\begin{figure}
\centerline{\psfig{figure=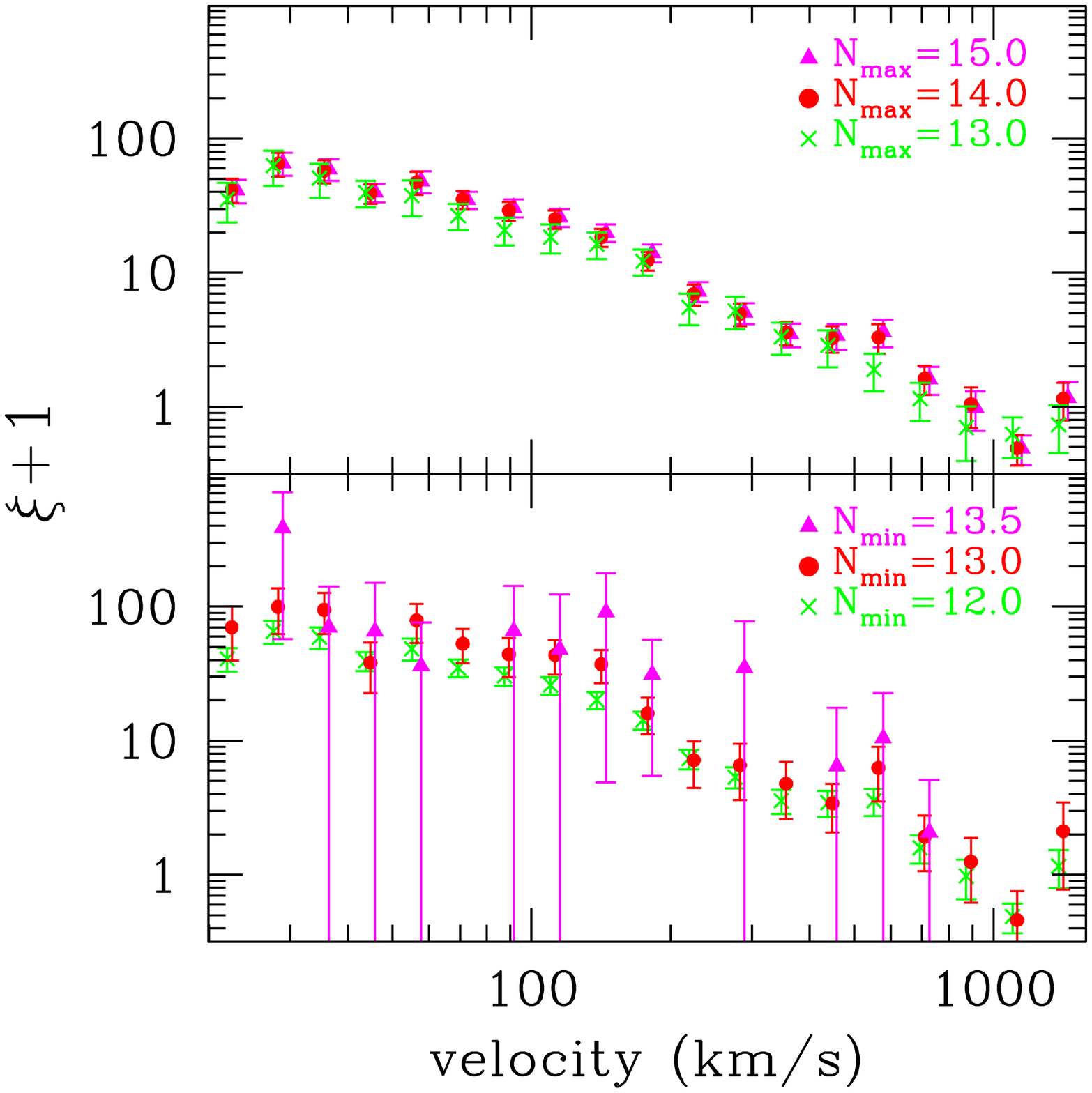,height=9.3cm}}
\caption{Dependence of the \CIV correlation function on column density
threshold.  {\em Upper panel:} Effect of applying a cut on the maximum
column density of \CIV components used to calculate $\xi_{\eCIV} (v)$.
In all cases $N_{\eCIV,\rm min} = 10^{12}$ cm$^{-2}$.
{\em Lower panel:} Effect of applying a cut on the minimum
\CIV column density, with $N_{\eCIV,\rm max}$ fixed at 10$^{16}$ cm$^{-2}.$}
\label{fig:CIVcuts}
\end{figure}

In Fig.\ \ref{fig:CIVcuts} we  study the dependence of the \CIV spatial
distribution on column density, by computing the correlation function over 
the full redshift range but selecting components within a fixed range of
column density.  In the upper panel of this figure we apply a cut on the
maximum column density component, while holding the minimum $N_\eCIV$
fixed at our detection limit of $10^{12}$ cm$^{-2}$.  
Apart from a weak shift in the 500-630 km s$^{-1}$ bin, 
$\xi_{\eCIV}(v)$ remains practically unchanged by this threshold.
As the majority of the detected components are relatively weak,
this indicates that our signal is determined by the bulk of the
components in our sample, rather than by properties of individual strong
absorbers.  

The results of a more drastic test are shown in the lower panel of
this figure.  Here we hold $N_{\eCIV, \rm max}$ fixed at $10^{16}$
cm$^{-2}$ and apply a cut on the minimum column density, which greatly
reduces the number of components in the sample.  Nevertheless moving
from $N_{\eCIV, \rm min} = 10^{12}$ cm$^{-2}$ to $N_{\eCIV, \rm min} =
10^{13.5}$ cm$^{-2}$ results only in a very weak enhancement of 
$\xi_{\eCIV}(v)$ at
small separations, while the rest of the correlation function
remains unchanged.  Thus, unlike \Lya absorption systems (Cristiani
\etal 1997), the correlation of \CIV does not depend strongly on
absorption column densities.  Instead, the spatial distribution seems
to be a global property of the population of \CIV components.

\begin{figure}
\centerline{\psfig{figure=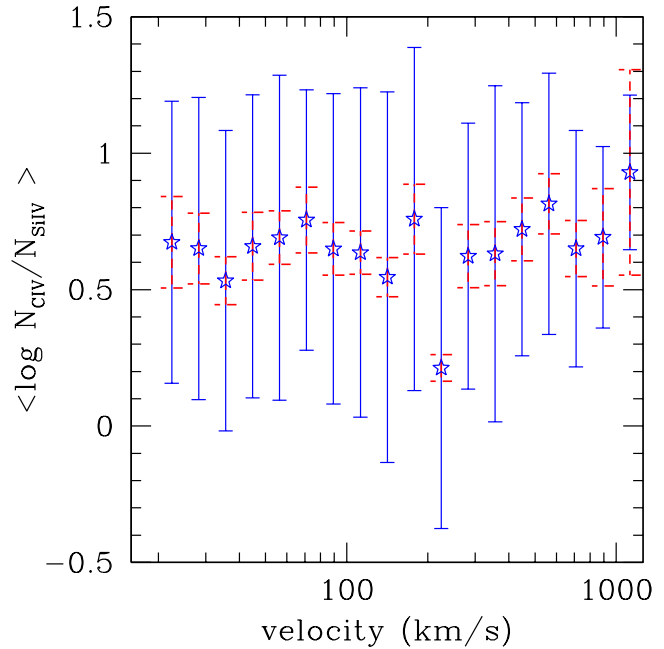,height=9.3cm}}
\caption{The average $\log(N_\eCIV/N_\eSiIV)$ ratio for
\CIV components contributing to the correlation function at various
separations.  At each separation the dashed error bars are the
statistical errors, while the solid error bars are the intrinsic
scatter.}
\label{fig:ratioseparation}
\end{figure}

A question that immediately arises is whether the features observed
in the \CIV and \SiIV correlation functions are intrinsic to the 
underlying  distribution of metals, or perhaps arise from variations 
in the UV background at somewhat shorter wavelengths.  
In fact, analyses of the \HeII distribution 
due to ionization by 54.4 eV photons
suggest that the second reionization of hydrogen may still have
been quite patchy at $z = 2.3- 2.9$ (Shull \etal 2004),
with \HeII found preferentially in ``void'' regions
where HI is weak or undetected.  

On the other hand, the ionization potentials of \CIII and \SiIII
are 47.5 eV and 33.5 eV, respectively, somewhat lower 
than that of {He{\sc ~ii}}, but well beyond the ionization potential of
hydrogen.  Thus if the suggested patchiness of \HeII
is due primarily to changes in the IGM opacity at wavelengths 
shortward of  54.4 eV, then the distribution of 
\CIV and \SiIV is likely to more closely trace the underlying
distribution of metals.  If \HeII inhomogeneities exist and are caused by 
a sparsity of hard sources, however, it is possible that background
variations may also play a role in the distribution of triply ionized
regions of carbon and silicon.

As the ionization potentials of \CIII and \SiIII differ by 12 eV, 
each is sensitive to a slightly different range of UV photons.  Thus
if the features seen in Figure \ref{fig:CIV} were produced by changes in the 
ionizing background, one might expect to see systematic changes in the 
ratio of these species as a function of separation.   As a simple test of 
this possibility, we considered the average $\log(N_{\eCIV}/N_{\eSiIV})$
as a function of separation.
In order to make the sample included in this average
as large as possible we computed this as
\ba
\left< \log (\frac{N_{\eCIV}}{N_{\eSiIV}}) \right>_k 
\hskip -0.2cm&=  \qquad \qquad \qquad \qquad \qquad \qquad \qquad \qquad 
\qquad  \nonumber
\\   & \sum_{i, j \in \tmop{bin}_k}
   \sum_{\ell}
\log \left( 
   \frac{N_{\eCIV,i}}{N_{\eSiIV,l}} \right)  \theta (5 -
   |v_{\ell} - v_i |)  \nonumber \\
 & \hskip -0.3cm \left\{ \sum_{i, j \in \tmop{bin}_k}
   \sum_{\ell} \Big[ 1  \times \theta (5 -
   |v_{\ell} - v_i | ) \Big] \right\}^{-1},
\ea
where $\theta(v)$ is the Heaveside step function,
$i$ and $j$ are indexes of \CIV components, 
$l$ is an index over all \SiIV
components, and $k$ is a given bin in velocity separation used to 
calculate the  correlation function.  In other words, for each bin in the correlation 
function,
we average $\log(N^{\eCIV}/N^{\eSiIV})$ over all \CIV components $i$ that are
found at the appropriate separation from another \CIV component $j$ and 
within 5 km s$^{-1}$ of a \SiIV component $l$.

The results of this analysis are found in Figure \ref{fig:ratioseparation},
which shows no correlation between separation and species abundances. 
Furthermore, our average value of
$\log(N^{\eCIV}/N^{\eSiIV}) \approx 0.7$ is similar to that seen
in previous analyses of 
$10^{12}$ cm$^{-2} \leq N_{\eCIV} \leq 3 \times 10^{14}$ cm$^{-2}$
absorbers (Kim, Cristiani, \& D'odorico 2002; BSR03), as well
as the weaker \CIV and \SiIV lines detected by Aguirre \etal (2003)
using the pixel optical depth method.
Thus there is nothing particularly unusual about the subset of
absorbers selected by our procedure.
Although this is clearly not an exhaustive test, it nevertheless suggests 
that the features in the correlation functions are not imprinted 
in a straightforward way by the UV background itself, and are more
likely to be caused by the spatial distribution of metals. 
However, a much more detailed analysis is necessary to settle this 
issue definitively.

\subsection{Peculiar Systems at Low-Redshift}

The \CIV correlation functions in Figs.\ \ref{fig:CIV} and \ref{fig:CIVcuts}
hint at a secondary bump at large separations.
It is important to try and understand if this 
comes from the presence of few peculiar systems or if this is
a generic feature of the \CIV distribution.  To this end
we computed the correlation function for different samples,
each time excluding one of the lines of sight, and
discovered that the signal comes from three QSOs, namely, 
PKS~0237$-$23, HE~0001$-$2340 and Q~0122$-$380.

\begin{figure}
\centerline{\psfig{figure=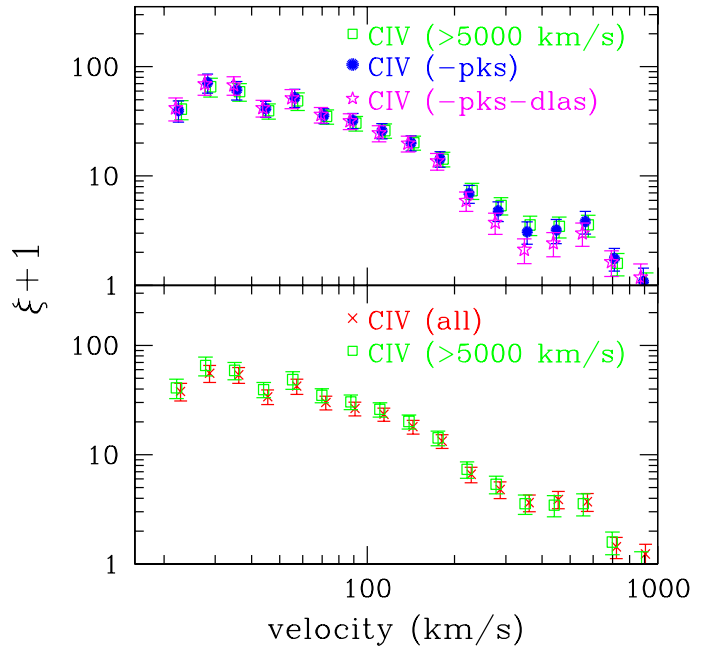,height=9.5cm}}
\caption{{\em Top:} Impact of peculiar systems on the 
\CIV correlation function.
Here the square points are computed from the full sample, the circles
are computed excluding the sightline towards PKS~0237$-$23,
and the stars are computed excluding both PKS~0237$-$23 and the
two sub-DLA systems, as described in the text.  {\em Bottom:}
Comparison between $\xi_{\eCIV}$ for the full sample, including 
associated absorbers (crosses), and excluding \CIV components
with 5000 km s$^{-1}$ of the quasar redshifts (squares).}
\label{fig:peculiar}
\end{figure}

The first of these has been long known to be very peculiar.
Indeed, a huge C~{\sc iv} complex is seen toward PKS~0237$-$23 at
eleven different redshifts over the range 1.596$-$1.676 (more than 
10,000~km~s$^{-1}$) 
with three main subcomplexes at $z_{\rm abs}$~=~1.596, 1.657 and 1.674 
(Boroson et al. 1978, Sargent et al. 1988). Furthermore Folz \etal (1993)
searched the  field around PKS~0237$-$23 
for other QSOs to provide background sources against 
which the presence of absorption at the same redshifts could be 
investigated. They concluded that the 
complex can be interpreted as a real spatial overdensity of absorbing 
clouds with a transverse size comparable to its extent along the line 
of sight, that is of the order of 30~Mpc. The correlation function without
this line of sight is shown in Fig.\ \ref{fig:peculiar}.

Two other lines of sight display peculiar systems. At 
$z_{\rm abs}$~=~2.1851 toward HE~0001$-$2340 there is a sub-DLA system and
the associated C~{\sc iv} system is spread over $\approx$450~km~s$^{-1}$.
It is therefore difficult to know if the structure there is due to large
scales or more probably to the internal structure of the halo associated 
with this high density peak. At $z_{\rm abs}$~=~1.9743 toward Q~0122$-$380, 
there is a double strong system spread over more than 500 km~s$^{-1}$. It
is again difficult to know whether these absorptions reflect internal
motions of highly disturbed gas.

After these are removed, the most significant excess at large
separations is found in the 500-630 km s$^{-1}$ bin.  This velocity
difference corresponds to the difference in wavelengths of the \CIV
doublet itself.  In fact, it is
interesting to note that this bin is the only one that is
significantly reduced by applying a cut to eliminate the larger
$N_{\eCIV}$ components, as was seen in Fig. \ref{fig:CIVcuts}.

As a further test of large-separation correlations, we have also
computed $\xi_{\eCIV}(v)$ including the associated systems, found
within 5000 km s$^{-1}$ of the redshifts of the QSOs in this sample.  
This is compared with the \CIV  correlation function for our standard
sample in the lower panel  of Fig.\ \ref{fig:peculiar}.  At all
separations, $\xi_{\eCIV}$ remains unchanged, thus indicating that
associated systems are not distributed in a particularly unusual way,
and do not contribute any significant features to $\xi_{\eCIV}(v)$  at
$\approx 500$ km s$^{-1}$, or any other separation.

\subsection{Iron II and Magnesium II}

\begin{figure}
\centerline{\psfig{figure=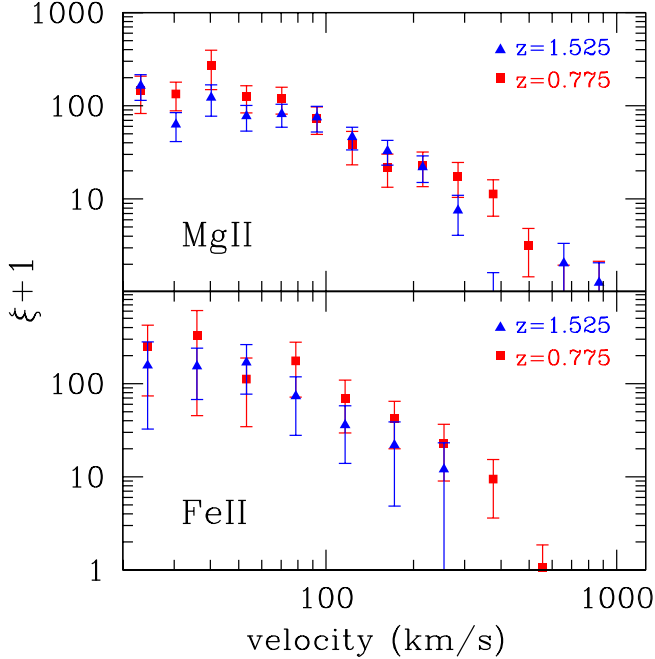,height=9.5cm}}
\caption{{\em Upper panel:} Two point correlation function of 
{Mg{\sc ~ii}},
divided into two redshift bins as in Figure 2. {\em Lower panel:} Two
point correlation function of {Fe{\sc ~ii}}, divided into the same redshift
bins.}
\label{fig:MgIIcomp}
\end{figure}

We now turn our attention to the distribution of lower-redshift
metals, as traced by \MgII and {Fe{\sc ~ii}}.   Splitting the data
into two redshift ranges yields the
line-of-sight  correlation functions shown in Fig.\
\ref{fig:MgIIcomp}, where again we have included both the Poisson and
sample-size errors in our estimate of the variances.  Like their
high-redshift counterparts, \MgII and \FeII are found to trace each
other closely. Their correlation functions are both relatively shallow  
at small separations and fall off more steeply at large separations.
Also like $\xi_{\eCIV}(v)$,  both $\xi_{\eMgII}(v)$ and
$\xi_{\eFeII}(v)$  exhibit slight enhancements at lower
redshifts, although again these excesses fall within the errors.

\begin{figure}
\centerline{\psfig{figure=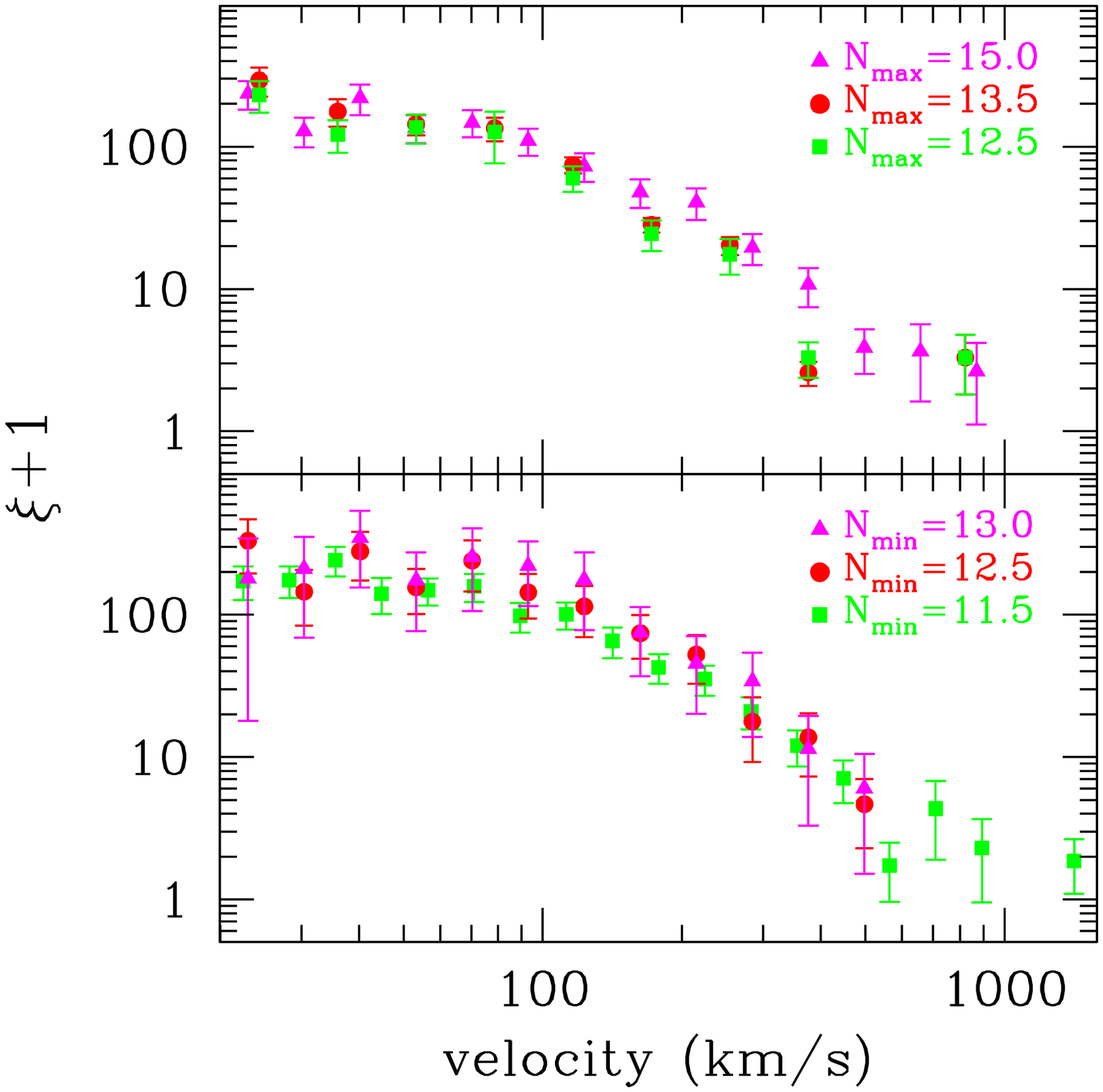,height=9.5cm}}
\caption{Dependence of \MgII correlation function on column density
threshold.  {\em Upper panel:} Effect of applying a cut on the maximum
column density of \MgII subcomponents used to calculate $\xi_{\eMgII}(v)$.  In
all cases $N_{\eMgII,\rm min} = 10^{11.5}$ cm$^{-2}$.  {\em Lower
panel:} Effect of applying a cut on the minimum \MgII column density,
with $N_{\eMgII,\rm max}$ fixed at 10$^{16}$ cm$^{-2}.$}
\label{fig:MgIIcuts}
\end{figure}

Next we examine the dependence of the \MgII spatial distribution on
column density.  Removing the strongest absorbers in our sample before
calculating $\xi_{\eMgII}(v)$ results in the values plotted in the upper panel
of Figure  \ref{fig:MgIIcuts}.  As in the \CIV case,
the \MgII correlation function is not dominated by the  clustering of
large components, but rather remains almost unchanged as a function of
$N_{\rm max}$, even when it  is reduced to $10^{12.5}$ cm$^{-2},$
excluding over a third of the systems.  Similarly, raising the minimum
column density from $10^{11.5}$ cm$^{-2}$ to $10^{12.5}$ cm$^{-2}$ does not
boost $\xi_{\eMgII}(v)$, even though this excludes $\approx 2/3$ of
the  sample.

\begin{figure}
\centerline{\psfig{figure=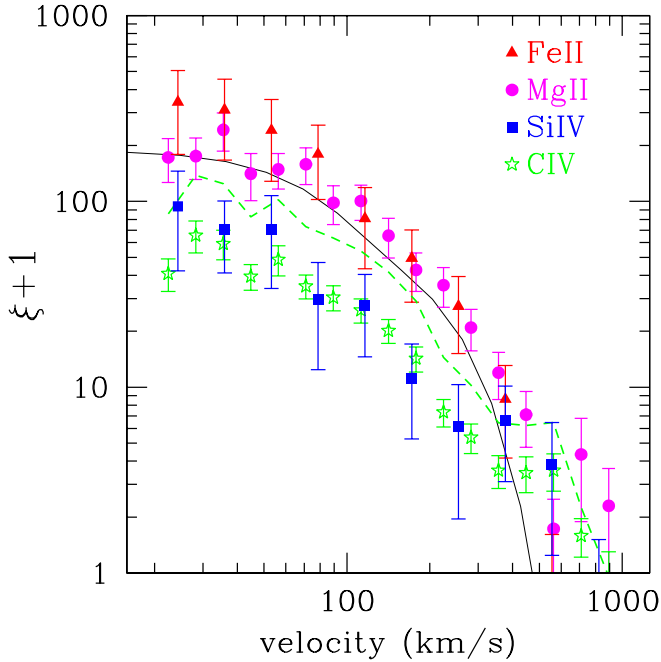,height=9.5cm}}
\caption{Measured correlation function of all metal-line components.
Points are measurements from our sample, while the solid line is the
\MgII fit by Churchill, Vogt, \& Charlton (2003), arbitrarily
normalized.  The dashed line is the \CIV correlation function, shifted
upwards by a factor of 2.1 to provide a simple estimate of the impact of
structure formation from $z=2.2$ to 1.1 on a fixed population of
absorbers.}
\label{fig:corrall}
\end{figure}

In Figure \ref{fig:corrall} we compare the correlation functions of
\CIV and \SiIV with those of \MgII and {Fe{\sc ~ii}}.  Note that the mean
redshift of these lower ionization species is $\approx 1.2$, while
for \CIV and \SiIV $z_{\rm mean} \approx 2.3.$  Thus our sample contains
very few objects in which all four species can be directly compared.
Nevertheless, a comparison of their redshift-space correlation
functions reveals a number of important parallels.
While $\xi(v)$ of all species decline steeply at large
separations and exhibit a turn-over at smaller velocity differences,
the transition between these two regimes is  pushed to slightly
smaller  separations in the \MgII and \FeII case, and the  fall-off at
higher  densities is more abrupt.

Interestingly, the features seen in this distribution can be inferred
from the original fitting to the distribution of velocity  separations
of \MgII absorbers by Petitjean \& Bergeron (1990), using a
remarkably small number of systems. Their data were fit with
the sum of two Gaussian distributions with similar overall weights
and velocities dispersions of $\sigma_v = 80$ km s$^{-1}$ and
$\sigma_v = $ 390 km s$^{-1}$, which the authors interpreted as due to the
kinematics of clouds bound  within a given galaxy halo, and the
kinematics of galaxies pairs, respectively.  Working at higher
spectral resolution and signal-to-noise,  Churchill, Vogt, \& Charlton
(2003) also obtained a good two-component  Gaussian fit to the
two-point clustering function of \MgII  components, although they did
not attempt to normalize this function to obtain $\xi_\eMgII (v) +
1$. In this case the best-fit values were $\sigma_v = 54$ km s$^{-1}$
and $\sigma_v = $ 166 km s$^{-1}$, where the relative amplitude of the narrow
component was twice that of the broad component.   This fit has been
added to  Figure \ref{fig:corrall}, adopting  an arbitrary
normalization.  Although our dataset has an overall signal-to-noise 
that is higher than that of  Churchill, Vogt, \&
Charlton (2003), and thus is more complete at  lower column densities,
their two-Gaussian model also provide a good
match to our data at $\Delta v \lesssim 400$ km/s.  However, it falls short
of the observed correlation at larger separations.

To contrast the correlation functions in more detail, we have
added a simple estimate of ``passive'' evolution 
to Figure \ref{fig:corrall}, that is the evolution if the metals detected at
$z \approx 2.3$ as \CIV absorbers were to move along with the formation
of structure before appearing as \MgII absorbers at $z=1.2$.   To
first approximation, the overall bias of such a metal tracer  field
would remained fixed, but its correlation function would be enhanced
by a factor of $D^2(1.2)$/$D^2(2.3) = 2.1$, where $D(z)$ is the
linear growth factor.  Surprisingly, simply shifting $\xi_{\eCIV}$ by
a factor of 2.3 provides us with an accurate match for the \MgII
correlation function over a large range of separations,
although it underpredicts the clustering of \MgII and \FeII at
smaller distances.  This is dicussed in futher detail in \S 8. 
%Thus it appears that all four species are approximately 
%tracing out the same (highly biased) cosmological
%regions.

To facilitate future comparisons, in Table 3 we give the correlation
function and errors for each of the four species averaged over our
full sample.   Note that the small number of \SiIV and \FeII
components forces us to use a smaller number of bins to beat down the
statistical noise in our measurements.

\begin{table*}
\centerline{
\begin{tabular}{lllllll}
\hline bin (km s$^{-1}$) & $\xi_\eCIV$ & $\xi_\eMgII$   & \qquad & bin (km s$^{-1}$) &
$\xi_\eSiIV$ & $\xi_\eFeII$ \\ \hline \\ 20-25      & 41    $\pm$ 8 &
170  $\pm$ 50 & \qquad  & 20-30    &   94  $\pm$ 52   &  310   $\pm$
150  \\ 25-32      & 66    $\pm$ 13     & 170  $\pm$ 40 & \qquad  &
30-43    &   71  $\pm$ 30   &  280   $\pm$ 130  \\ 32-40      & 59
$\pm$ 11     & 240  $\pm$ 60 & \qquad  & 43-65    &   71  $\pm$ 36   &
220   $\pm$ 100  \\ 40-50      & 40    $\pm$ 6      & 140  $\pm$ 40 &
\qquad  & 65-100   &   30  $\pm$  17  &  160   $\pm$ 70   \\ 50-63 &
49    $\pm$ 9      & 145  $\pm$ 30 & \qquad  & 100-140  &   27 $\pm$
13  &  74    $\pm$ 34   \\ 63-79      & 35    $\pm$ 5      & 155
$\pm$ 40 & \qquad  & 140-200  &   11  $\pm$  6   &  45    $\pm$ 19
\\ 79-100     & 30    $\pm$ 5      & 96   $\pm$ 23 & \qquad  & 200-300
&   6.2 $\pm$  4.2 &  25    $\pm$ 11   \\ 100-125    & 26 $\pm$ 4
& 98   $\pm$ 21 & \qquad  & 300-450  &   6.6 $\pm$  3.5 & 7.8   $\pm$
4.1  \\ 125-160    & 20    $\pm$ 3      & 64   $\pm$ 15 & \qquad  &
450-670  &   3.9 $\pm$  2.6 &  0.88  $\pm$ 0.58  \\ 160-200 & 14
$\pm$ 2      & 42   $\pm$ 10 & \qquad  & 670-1000 &   0.8 $\pm$  0.7 &
0     $\pm$ 1     \\ 200-250    & 7.3   $\pm$ 1.2    & 35   $\pm$ 8  &
\qquad  & \qquad   & \qquad         & \qquad \\ 250-320    & 5.4
$\pm$ 1.0    & 20   $\pm$ 5  & \qquad  & \qquad   & \qquad         &
\qquad \\ 329-400    & 3.6   $\pm$ 0.7    & 12 $\pm$ 3  & \qquad  &
\qquad   & \qquad         & \qquad \\ 400-500 & 3.5   $\pm$ 0.7    &
6.9  $\pm$ 2.3 & \qquad & \qquad   & \qquad & \qquad \\ 500-630    &
3.6   $\pm$ 0.8    & 1.7  $\pm$ 0.8 & \qquad & \qquad   & \qquad
& \qquad \\ 630-790    & 1.6   $\pm$ 0.4 & 4.2  $\pm$ 2.4 & \qquad &
\qquad   & \qquad         & \qquad \\ 790-1000   & 0.98  $\pm$ 0.32
& 2.2  $\pm$ 1.3 & \qquad & \qquad   & \qquad         & \qquad \\
\hline\
\end{tabular}}
\label{xifunctions}
\caption{Summary of measured metal-component correlation functions.
Note that there is likely to be significant redshift evolution
of these functions. The mean redshifts  of \CIV and \SiIV are
$\approx 2.3$.  The mean redshifts of \MgII and \FeII are $\approx 1.15.$
}
\end{table*}

\begin{figure}
\centerline{\psfig{figure=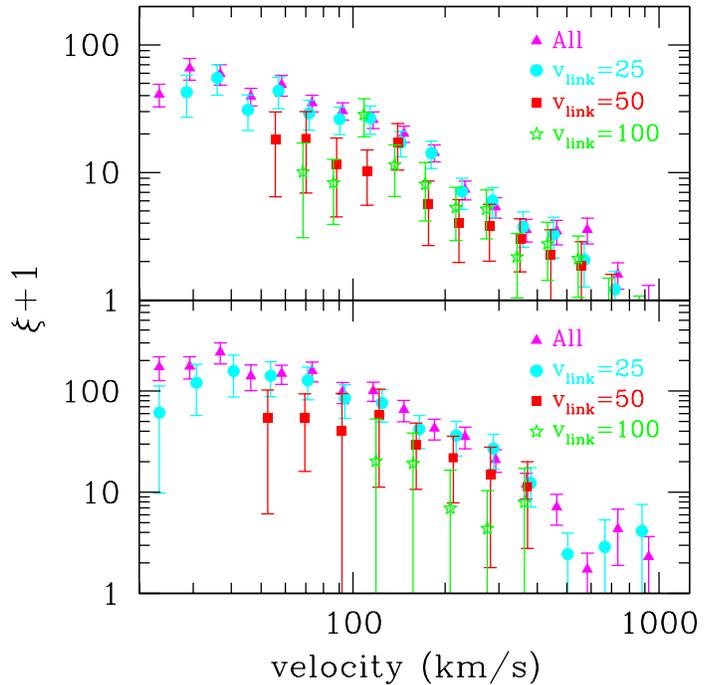,height=10cm}}
\caption{{\em Upper panel:} Impact of linking together  \CIV
components into systems.   {\em Lower panel:} Impact of linking
together  \MgII components into systems. Details described in the
text.}
\label{fig:bothlink}
\end{figure}

Finally, we carry out a test to determine if the spatial distribution
of metals as traced by $\xi_{\eCIV}(v)$ may be affected by our  VPFIT
decomposition into components.  Previous studies have attempted to trace
the distribution of intergalactic metals by grouping together
components into ``systems,'' which are likely to have a common
physical origin, and computing the correlation function of these
systems (e.g.\ Petitjean \& Bergeron 1990; BSR03).  While
typically, system identifications have been carried out by eye, here we
attempt a more objective approach, which parallels the
friends-of-friends technique (Davis \etal 1985)  widely used for
group-finding in cosmological simulations.  In this case, we define a
velocity linking-length ($v_{\rm link}$)  and group together all
components whose separation from their  nearest neighbor is
less than $v_{\rm link}$ into
a system at a redshift equal to the average over all its components.
Note that this procedure does not  involve simply linking together
pairs within $v_{\rm link},$ but rather forms collections of many
components, each within a linking length of its neighbors and grouped
together  into a single entity.    It is therefore equivalent to
partitioning a set of components into two  systems whenever they are
separated by a gap wider than $v_{\rm link}.$

In the upper panel of Figure \ref{fig:bothlink} we plot $\xi_{\eCIV}(v)$
computed for the resulting \CIV systems, for three different choices
of $v_{\rm link}$.  In all cases, within our  measurement errors,
combining components into systems has  no appreciable impact at
separations much larger than the  linking length. Thus while
BSR03 report a lack of clustering of systems as
identified by eye, we are unable to reproduce this behavior with our
automatic method.  Perhaps this is not surprising, as the clustering
of $\xi_\eCIV(v)$ is very strong, and thus many pairs of ``systems''
are likely to be closely spaced and easily tagged as a single object.  
However, our results show that fixing a pre-specified
definition of systems does not remove large-scale correlations in this
way.

In the lower panel of Figure \ref{fig:bothlink}, we see that grouping
\MgII  components into systems has no clear impact at larger
separations if $v_{\rm link} = 25,$  $50$, and while there are hints
of weak larger-scale damping if $v_{\rm link} = 100$, these changes
are within our errors.  Similar results were obtained if each group
was assigned the redshift of its largest component, leading us to
conclude that the $\xi(v)$ features observed in both the high-redshift
and low-redshift species are not related to division into
components, but rather reflect the underlying distribution of
intergalactic metals.

\section{Numerical Simulation}

\subsection{Overall Properties}

To better interpret the features seen in metal absorption-line
systems, we conducted a detailed smoothed particle hydrodynamic (SPH)
simulation of structure formation. Our goal here is to apply 
the same 
automated procedure used to identify metal absorbers in the LP data set
to a detailed simulation, drawing conclusions as to what constraints
our measurements place on the underlying distribution of IGM metals.
For this purpose we focus our attention on a
cold dark matter cosmological model  with the same general
cosmological parameters as above, and the additional parameters
$\sigma_8 = 0.87$, and $n=1$, where $\sigma_8^2$ is the variance of
linear fluctuations on the $8 h^{-1}{\rm Mpc}$ scale, and $n$ is the
``tilt'' of the primordial power spectrum.  The Bardeen et al.\ (1986)
transfer function was used with an effective shape parameter of
$\Gamma=0.2$, and the ionizing background flux was  taken to be
a Haardt \& Madau (1996) spectrum with 
$J(\nu,z) = 2.2 \times 10^{-22} \,
\left(\frac{\nu}{\nu_{\rm HI}}\right)^{-1}  \, (1+z)^{0.73}
\, \exp \left[ -(z-2.3)^2/1.9 \right]$ ergs s$^{-1}$ Hz$^{-1}$
cm$^{-2}$ sr$^{-1}$, corresponding to a photoionization rate of $6.8
\times 10^{-13}  \, (1+z)^{0.73} \, \exp \left[ -(z-2.3)^2/1.9 \right]$
s$^{-1}.$

We simulated a box of size $40/h$ comoving Mpc on a side, using
$320^3$ dark matter particles and an equal number of gas particles.
The mass of each dark matter particle was $2.0 \times 10^{8} \msun$,
and the mass of each gas particle was $3.4\times 10^{7} \msun$. This
yields a nominal minimum mass resolution for our (dark matter) group
finding of $1.0 \times 10^{10} \msun.$ 
The run was started at an initial redshift of
$z=99$, and a fixed physical S2 softening length of 6.7 kpc was
chosen, which is equivalent to a Plummer softening length of 2.8 kpc.
The simulation was conducted using a parallel MPI2-based version of
the HYDRA code (Couchman \etal 1995) developed by the Virgo Consortium
(Thacker \etal 2003).

We used the SPH algorithm described in Thacker \etal (2000),
although in an improvement upon earlier work, the maximum SPH search
radius now allows us to accurately resolve the mean density of the
box.   Photoionization was implemented using the publicly available
routines from our serial HYDRA code.
Radiative cooling was  calculated using the Sutherland \& Dopita
(1993) collisional ionization equilibrium tables, and a uniform 2\%
solar metallicity was assumed for all gas particles for cooling
purposes. Integration  to $z=2.0$ required 9635 (unequal) steps, and
four weeks of wall clock time on 64 processors.  Outputs for
post-processing were saved at redshifts of $z=8.0,$ 5.0, 4.0, 3.0,
2.5, and 2.0.

From each of the final three outputs, we interpolated to extract
two-dimensional slices of overdensity, temperature, and line-of-sight
peculiar velocities on 24 equally spaced planes.  By extracting random
sight-lines from each of these three fields, we were then able to
generate  simulated metal-line spectra, which could be processed in an
identical fashion as the observed data.  Before turning our attention
to this issue, however, we first address the more basic concern of the
overall hydrogen distribution, which serves as both a check of our
simulation methods, and a way of fine-tuning the assumed ionizing
background to reproduce the observed properties of the IGM.

\subsection{Calculation of Neutral Hydrogen Fraction}

Once the baryon density, temperature, and line-of sight velocity are
 extracted along a line of sight, constructing a simulated spectrum is
 relatively straightforward.  One obtains the  neutral hydrogen fraction, 
$f_{\rm HI}$, in the IGM by solving the ionization
 equilibrium equation (Black 1981) 
\be 
\alpha(T) n_p n_e=\Gamma_{\rm
 ci}(T) n_e n_{\rm HI} + J_{22} G_1 n_{\rm HI},
\label{eq:ion_eq}
\ee  where $\alpha(T)$ is the radiative recombination rate,
$\Gamma_{\rm ci}(T)$ is the rate of collisional ionization,  $J_{22}$
is the UV background intensity  in units of $10^{-22}$ ergs s$^{-1}$
Hz$^{-1}$ cm$^{-2}$ sr$^{-1}$,  $J_{22} G_1$ is the rate of
photoionization, and $n_p, n_e$ and $n_{\rm HI}$ are the number
densities of protons, electrons, and neutral hydrogen, respectively.
For the Haardt \& Madau (2001) spectrum assumed below $G_1 = 2.7
\times 10^{-13}$ s$^{-1}$, for the original Haardt \& Madau
(1996) spectrum $G_1 = 3.2 \times 10^{-13}$ s$^{-1}$, and for the
$(\nu/\nu_{\rm HI})^{-1}$ spectrum used in our simulation $G_1 = 3.1 \times
10^{-13}$ s$^{-1}$.  For comparison, $G_1 J_{22}$ is equal to $J_{-12}$
as defined in  Choudhury \etal (2001) and $10 G_1 J_{21}$ as defined
in Bi and Davidson (1997).

If we assume that the neutral fraction of hydrogen $\ll 1$ and all the
helium present is in the fully-ionized form, we find 
\be 
f_{\rm
HI}(x,z)=\frac{\alpha \left[ T(x,z) \right] } {\alpha
\left[T(x,z) \right]+\Gamma_{\rm ci}  \left[ T(x,z) \right] + G1 J_{22}
n_e^{-1}(z)},
\label{eq:nh_nb}
\ee 
where 
the recombination rate $\Gamma_{\rm ci}(T)= 5.85\times10^{-11} T^{1/2}
\exp(-157809.1/T)~{\rm cm}^3{\rm s}^{-1},$ with $T$ in Kelvin, and
Black (1981) gives an approximate form for
the recombination rate as  
\be 
\frac{\alpha(T)}{{\rm
cm}^3{\rm s}^{-1}}=\left\{ \begin{array}{ll}
4.36\times10^{-10}T^{-0.7573} &\mbox{(if $T \geq 5000$K)}\\
2.17\times10^{-10}T^{-0.6756} &\mbox{(if $T < 5000$K).}
		\end{array} \right.
\ee 

With these expressions we can compute the neutral hydrogen density,
$n_{\rm HI}[x,z(x)]$,  along a line of sight.  Here $x$ and $z$ are
related by $c \, {\rm d}z 
= {\rm d}x \, H(z)$, where the Hubble constant as a function
of redshift is $H(z) = H_0 \sqrt{\Omega_\Lambda +  \Omega_m (1+z)^3 }.$
We choose a coordinate system such that $x = 0$ at the front of the
box and define $\Delta z(x,z_0)$
as the change in redshift from $x=0$.
We then construct the \Lya optical depth as 
\ba 
\tau_\alpha(z_0+ \Delta z)  = & \hskip-1.2cm \displaystyle{
\frac{c \sigma _{\alpha}}
{(1+z_0) \sqrt{\pi}}  \int~{\rm d}x   \frac{n_{\rm
HI}(x,z_0)}{b(x,z_0)}} \times \\
 & \exp\left[-\left( \frac{x H(z_0) +
v(x,z_0)(1+z_0) - c \Delta z} {(1+z_0)b(x,z_0)}\right)^2\right],
\nonumber
\label{eq:tauHI}
\ea
where $b(x,z_0) \equiv \sqrt{\frac{2k_{\rm B}T(x,z_0)}{m_p}}$ (with
$k_B$ the Boltzmann constant),
$n_H(z) = 1.12 \times 10^{-5} (1-Y) \Omega_b (1+z)^3 h^2 \, {\rm
cm}^{-3} = 1.83 \times 10^{-7} (1+z)^3 \, {\rm cm}^{-3}$
(with $Y$ the helium mass fraction), and
$\sigma_\alpha$ is the \Lya cross section, which can be calculated as
\be 
\sigma_\alpha = (3 \pi \sigma_T/8)^{1/2} f \lambda_0,
\label{eq:Xsec}
\ee where $\lambda_0$ is the restframe wavelength of the transition,
$f$ is the appropriate 
oscillator strength, and  $\sigma_T = 6.25 \times 10^{-25}$
is the Thomson cross section.   For Ly$\alpha$, $\lambda_0 =1215$ \AA
\, and $f = 0.4162$, which gives $\sigma_\alpha = 4.45 \times
10^{-18} {\rm cm}.$
With eq.\ (\ref{eq:tauHI}) in hand, we are able to construct simulated
UVES spectra of the Lyman-$\alpha$ forest by stacking vectors of
optical density computed from randomly extracted sightlines.  These are
then convolved with a Gaussian  smoothing kernel with a width of 4.4
km s$^{-1}$ (corresponding to the UVES resolution) and rebinned onto a
205,000 array of wavelengths, using the UVES pixelization from 3,050
to 10,430 \AA.  Rather than interpolate between simulation outputs,
however, we first turn our attention to a careful comparison between
observations and limited segments of spectra 
at fixed redshifts, concentrating on two main quantities: the
probability distribution function, a single-point quantity that is
sensitive to the overall temperature and $J_{22}$ evolution
background, and the two-point correlation function, a measure of the
spatial distribution of the gas.

\begin{figure}
\centerline{\psfig{figure=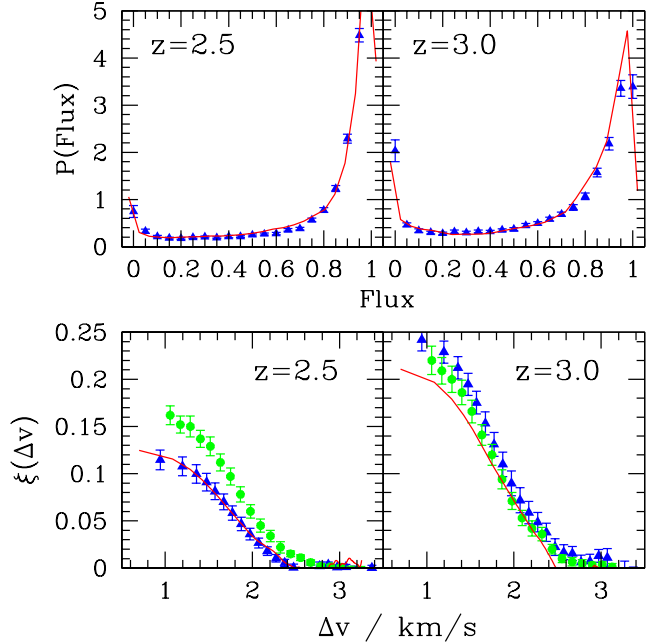,height=9.5cm}}
\caption{{\em Top:} Measured and simulated flux PDFs of the Lyman-$\alpha$
forest at two representative
redshifts.  Measurements are taken from McDonald \etal (2000) over
redshift ranges of $2.09\leq z \leq 2.67$ (left panel) and $2.67 \leq
z \leq 3.39 $ (right panel), respectively, while simulations are at
fixed redshifts of 2.5 (left panel) and 3 (right panel).
{\em Bottom:} Measured and simulated flux two-point correlation functions
 of the Lyman-$\alpha$ forest;
$\xi(\Delta v) = < \delta F(v) \delta F(v+\Delta v)>$
where $\delta F = F/\bar F - 1$, at two representative
redshifts.  The triangles are 
measurements by McDonald \etal (2000)
over redshift ranges of $2.09\leq z \leq
2.67$ (left panel) and $2.67 \leq z \leq 3.39 $ (right panel)
respectively and the circles are measurements by Croft \etal (2002) 
over redshift ranges of $2.31\leq z \leq
2.62$ (left panel) and $2.88 \leq z \leq 3.25 $ (right panel).
Again the simulations, represented by the solid lines,
are at fixed redshifts of 2.5 (left panel) and 3 (right panel).}
\label{fig:PDF}
\end{figure}

\subsection{Tests of The Numerical Hydrogen Distribution}

The probability distribution (PDF) of the transmitted flux was first
used to study the \Lya forest by Jenkins \& Ostriker (1991) and since
then has been a widely used tool for quantifying the mean properties
of the IGM (e.g.\ Rauch \etal 1997, McDonald \etal 2000). 
In the upper panels of 
Figure \ref{fig:PDF} we compare the PDF as measured by McDonald
\etal  (2000) to that generated from 20 simulated spectra at
representative redshifts of 2.5 and 3.0.  In order to obtain the good
agreement seen in  this figure, it was necessary to adjust the assumed
$J_{22}$ values to 4.7 at $z$ = 2.5 and 3.7 at $z$ = 3 (corresponding
to photoionization rates of $1.3 \times 10^{-12}$ s$^{-1}$ and  $1.0
\times 10^{-12}$ s$^{-1}$), down from the values of 5.4 and 4.7
(corresponding to photoionization rates of  $1.7 \times 10^{-12}$ s$^{-1}$ 
and $1.5 \times 10^{-12}$ s$^{-1}$) respectively, that were assumed in the
simulations.

This results in a slight inconsistency between the simulated
$\rho$-$T$ relation and the one that would have arisen if we had
repeated the simulation with our fit values of $J_{22}$.  In practice
however, this difference is unimportant in relation to our primary goal
of constructing simulated metal lines.  It is dwarfed by effects
due  to the uncertain evolution of the UV background at higher
redshifts (e.g.\  Hui \& Gnedin 1997; Hui  \& Haiman 2003),
uncertainties in the normalization of the quasar spectra
(e.g.\ McDonald 2000) and the extrapolation of the UV background from
912\AA \, to the shorter wavelengths relevant for \CIV and \SiIV (\eg
Haardt \& Madau 2001).  Thus our approach is more than adequate for
the purposes of this study.  With our assumed background values, the
mean fluxes at $z=2.5$ and $3.0$ are 0.794 and 0.692 respectively,
while the observed values are $0.818 \pm 0.012$ and $0.684 \pm 0.023$.

As a second test of our simulations, we constructed the Lyman-$\alpha$
flux correlation function $\xi(\Delta v) =  < \delta F(v) 
\delta F(v+\Delta v)>,$ which primarily provides a validation of our
assumed primordial power spectrum $P(k)$ and its evolution in our
simulation.   Beginning with Croft \etal (1998), the direct
inversion of the one dimensional power-spectrum of the Ly$\alpha$ flux
has been seen as one of the best constraints on the shape of the mass
power spectrum on intermediate scales (\eg Hui 1999; 
McDonald \etal 2000; Pichon \etal 2001; Croft \etal 2002; 
Viel, Haehnelt, \& Springel 2004).

Again, this quantity is straightforward to extract from our simulated
spectra.  The resulting values are shown in the lower panels
of  Figure \ref{fig:PDF},
in which we compare them to measurements by McDonald \etal  (2000) as
well as  Croft \etal (2002).  As in the single-point case, our
simulations are generally in good agreement with the observed values.
In fact, at $z = 2.5$, our simulated values are well within the range
of values bracketed by the weakly disagreeing observational results.
At $z= 3.0$,  our simulated values provide a slight underprediction
at small separations, although this is only just outside the 1-$\sigma$
error in the current measurements.  In summary then, the gas
properties of our numerical simulation are more than adequate to
provide a firm basis for the construction of Ly-$\alpha$ spectra,
while  at the same time containing sufficient resolution to allow us
to push  toward the denser regions associated with metal-line
absorption systems.

\begin{figure}
\centerline{\psfig{figure=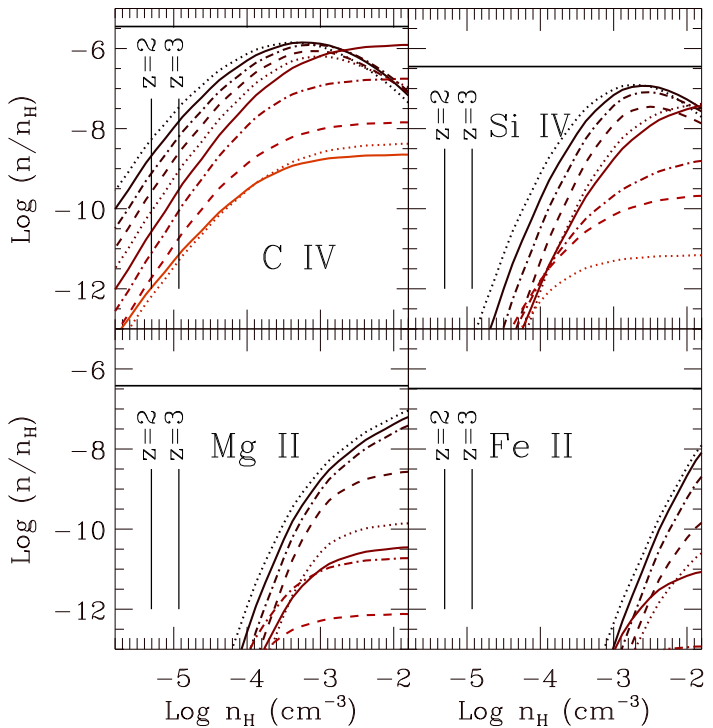,height=10.5cm}}
\caption{Abundances of various species as a function of total hydrogen
number density for a $10^{-2} \, Z_\odot$ gas exposed to a Haardt \&
Madau (2001) background at $z=2.5$ In the upper left panel, the
temperatures corresponding to each of the curves are, from top to
bottom, $10^{3.75}$ K (dotted), $10^{4}$ K (solid), $10^{4.25}$ K
(dot-dashed), $10^{4.5}$ K (dashed), $10^{4.75}$ K (dotted),
$10^{5.0}$ K (solid), $10^{5.25}$ K (dot-dashed), $10^{5.5}$ K
(dashed), $10^{5.75}$ K (dotted), and $10^{6.0}$K (solid).  Similar
labeling is used in the other panels.  The vertical lines give the
mean hydrogen number density at $z=3$ and 2 while the horizontal lines
give the total abundance of each of the elements.}
\label{fig:cloudy1}
\end{figure}

%\begin{figure}
%\centerline{\psfig{figure=cloudy2.eps,height=12cm}}
%\caption{Number density of various species relative to 
%{H{\sc ~i}}. Curves
%are as in Figure \protect{\ref{fig:cloudy1}}}.
%\label{fig:cloudy2}
%\end{figure}

\begin{figure*}
\begin{center}
\resizebox{11.1cm}{!}{\includegraphics{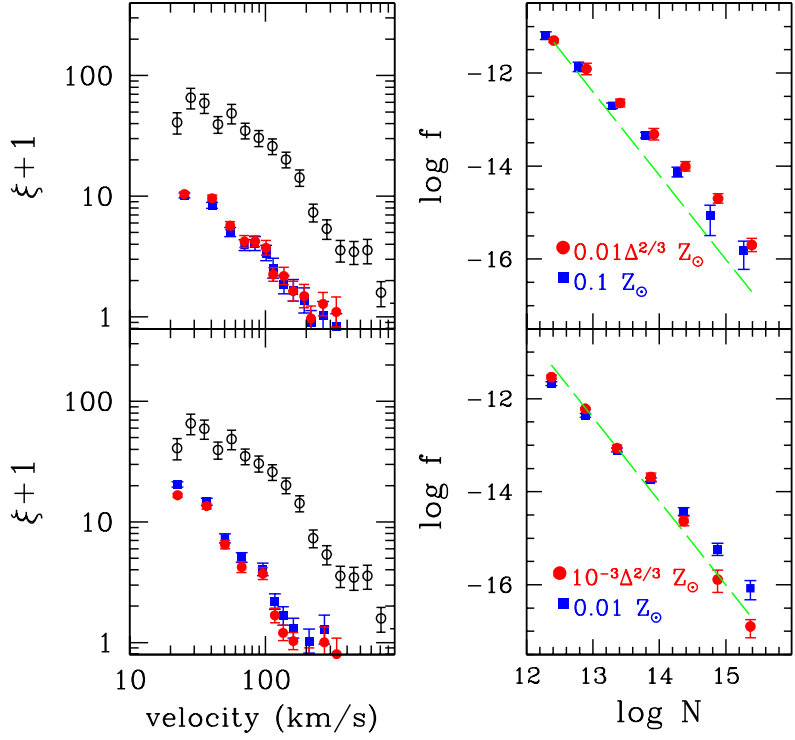}}\\%
\caption{Correlation functions and column density distributions for
models in which metallicity is assumed to be constant throughout the
simulation, or a simple function of density.  The filled points in the
upper panels show high-metallicity models in which $Z = 10^{-1}
Z_\odot$ and $Z = \Delta^{2/3} 10^{-2} Z_\odot$, while the filled
points in the lower panels show lower metallicity models in which  $Z
= 10^{-2} Z_\odot$ and $Z= \Delta^{2/3} 10^{-3} Z_\odot.$ Fifty-seven
simulated QSO sightlines were averaged in the high-metallicity models
and 114 were averaged in the lower-metallicity cases.  The open
circles give the measured \CIV correlation function, and the dashed
lines give the fit to the column density distribution as in Figure
\ref{fig:abundance}.}
\label{fig:joop}
\end{center}
\end{figure*}

\section{Modeling Metal Enrichment}

\subsection{Calculation of Observed Metal Lines}

Extending the methods of \S5.2 to construct the spatial
distribution of metal lines requires us to adopt an overall spectral
shape for the ionizing background, as well as a more detailed
calculation of the densities of various species.  Here we assume a UV
spectrum as predicted by the  updated models of Haardt \& Madau (2001;
see also Haardt \& Madau 1996) at $z=2.5$, but shifted such that
$J_{22}$ is consistent with the levels found in
$\S5.$ Assuming local thermal equilibrium, we then make use of
CLOUDY94 (Ferland \etal 1998; Ferland 2000) to construct tables 
of the relevant species as a function of temperature and
 density at each of these redshifts, for a characteristic metallicity
of $10^{-2} \zsun$.  Self-shielding in optically
thick regions was not taken to account.  The resulting species densities 
are shown as a function of hydrogen
number density and temperature 
in Fig.\ \ref{fig:cloudy1}, which is modeled after
Figure 2 of Rauch, Haehenelt, \& Steinmetz (1997).

In this figure, we see that, roughly speaking, \CIV traces the widest
range  of environments, while {Si{\sc ~iv}}, {Mg{\sc ~ii}}, and \FeII
probe progressively denser regions.  Thus while an appreciable level
of \CIV is found in only a few times overdense $z=3$ gas, comparable
levels of \SiIV are achieved only in 
denser regions with $\Delta \equiv \rho(x)/\bar \rho \approx 10$; and
while \MgII is found at similar densities to {Si{\sc ~iv}}, \FeII is only
detectable in $\Delta \ge 100$ regions, orders of magnitude 
denser than most \CIV regions.

Similarly, \CIV is detectable over a large temperature range, covering
from 10$^4$ K up to $\approx 10^6$ K. While \SiIV is also relatively
stable with respect to temperature changes, \MgII and \FeII are much
more fragile, and their abundances fall away quickly above $\approx 10^5$
K.  From these results, we see that  the correct modeling of \SiIV
requires  simulations  that probe to densities $\approx 10$ times higher than
those most  relevant to C{\sc ~iv},
although $\xi_{\eCIV}(v)$ and
$\xi_{\eSiIV}(v)$ trace each other closely.  Thus while our numerical
modeling was carried out at the highest resolution possible, we
nevertheless limit our comparisons  to \CIV to minimize any remaining
numerical effects.

\subsection{A Nonlocal Dependence}

Having determined the number densities of each of the species of
interest as a function of temperature and density in a $10^{-2} \zsun$
medium, we then applied these calculations to extract simulated metal
absorption spectra from our simulations. As a first step, 
following Rauch, Haehnelt, \& Steinmetz (1997), we 
assumed a constant metallicity across the simulation volume
and extracted sightlines of $\tau_{\eCIV}$ using an appropriately
modified version of eq.\ (\ref{eq:tauHI}): 
\ba
\tau_{\eCIV,i}(z_0+
\Delta z) =  & \hskip-1.2cm \displaystyle{\frac{c 
\sigma _{\eCIV,i}}{(1+z_0) \sqrt{\pi}}  \int~{\rm
d}x   \frac{n_{\eCIV}(x,z_0)}{b_{\eCIV}(x,z_0)} \times} \\
& \exp\left[-\left(
\frac{x H(z_0) + v(x,z_0)(1+z_0) - c \Delta z}{(1+z_0)
b_{\eCIV}(x,z_0)}\right)^2\right], \nonumber
\ea 
where now  $b_{\eCIV}(x,z_0)_
\equiv  \sqrt{\frac{2k_{\rm B}T(x,z_0)}{12 m_p}},$
$n_{\eCIV}(z)$ is the mean \CIV density, and
$\sigma_{\eCIV,i}$ is the cross section  corresponding to the $i$th
absorption line of the \CIV  doublet.  These we compute directly from
eq.\ (\ref{eq:Xsec}) taking $(\lambda_{0,\eCIV,1},\lambda_{0,\eCIV,2})
= (1548.2,1550.8)$ and $(f_{\eCIV,1},f_{\eCIV,2}) =
(0.1908,0.09522)$.  For the low metallicities relevant for the IGM,
the effects of changing metallicity can be modeled as a simple linear
shift in the species under consideration.

In contrast with the fixed-redshift comparisons described in \S5, our
goal was to construct simulated data sets that corresponded as closely
as possible to the full ESO Large Program data set.  In this case,
instead of stacking together spectra drawn from a single output,
we instead allowed for redshift evolution:
drawing slices from the output that most closely corresponded to the
redshift in question,  taking $n \propto (1+z)^3$,  and interpolating
between CLOUDY tables with appropriate $J_{22}$ values.  Finally,
we applied Poisson noise corresponding to  a signal-to-noise of 100 per 
pixel.

Each spectrum generated by this method was subject to 
the same two-step identification procedure that was applied to the
real data, and the resulting fits were subject to the same $N$ and $b$
cuts described in \S2.2.  The line lists compiled in this way were
then used to generate correlation functions and column density
distributions that directly parallel those calculated from the Large
Program data set.

These are shown in Fig.\ 13, in which we explore a 
low-metallicity model ($10^{-2} Z_\odot$) roughly consistent with previous
estimates (e.g.\ Rauch, Haehnelt, \& Steinmetz 1997; Schaye \etal 2003)
as well a higher-metallicity model ($10^{-1} Z_\odot$).   Note that at
these metallicities, changes in $Z$ can be modeled  simply by boosting
the \CIV density derived from the CLOUDY tables by a linear
factor. Increasing the metallicity by a factor of 10 in  this way has
very little effect on  the correlation function: decreasing
$\xi_{\eCIV}(v)$ in  the  20 and 35 km s$^{-1}$  bins by roughly a factor of
1.5, while leaving the rest of the correlation function largely
unchanged. In all  cases these values fall far short of the clustering
levels seen in our observational data, and they lack the conspicuous bend
observed at $\approx 150$ km s$^{-1}$.

\begin{figure*}
\begin{center}
\resizebox{11.9cm}{!}{\includegraphics{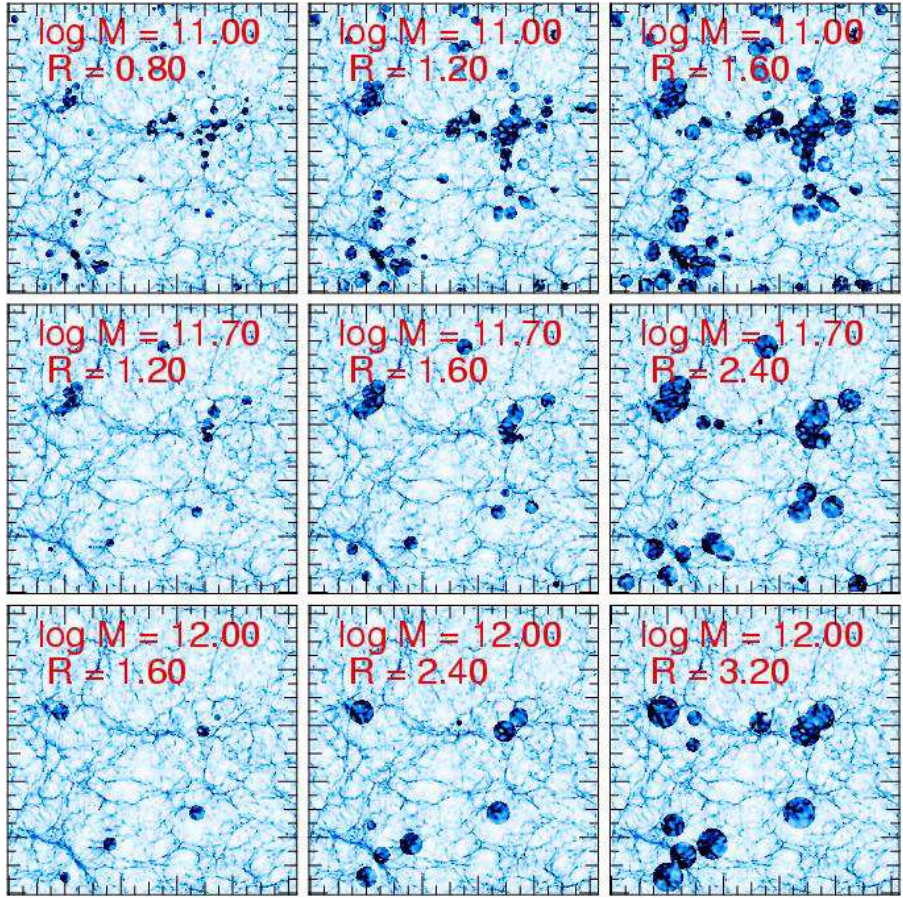}}\\%
\caption{The \CIV distribution of a $z=3$ slice in the simulation,
dark regions are those contained within a distance $R_s$ from a  dark
matter halo of mass $M_s$, with parameters as labeled in the panels.
These regions are used in constructing simulated spectra, while all gas
outside them is considered to be metal-free.  From left to right, and
then top to bottom, the overall volume filling factors of these models
are:  3.3\%, 8.6\%, 16.1\%;  3.4\%, 8.6\%, 16.7\%; and 1.7\%, 5.5\%,
11.6\%}
\end{center}
\label{fig:paint}
\end{figure*}

However, changing metallicity has a large effect on the column
density distribution.  The low-metallicity model is consistent with
observations over the range of  $12.5 \leq \log N   [{\rm cm}^{-2}]
\leq 13.5$,  and slightly overpredicts the number of large components
(which have a negligible impact on  the correlation function).  The
high-metallicity model,  on the other hand, overpredicts the number of
components for all column densities $\log N [{\rm cm}^{-2}]\ge 13.0.$

The poor fit to the correlation function is perhaps not surprising
given the known inhomogeneity of the IGM metal distribution (\eg
Rauch, Haehnelt, \& Steinmetz 1997).   Most recently, this has been
quantified by Schaye \etal (2003), who applied a pixel optical depth
method to derive a nonlinear relation between the local overdensity
$\Delta$ of hydrogen and the local carbon
abundance.   Over a large range of environments, they found $[C/H]
\propto \Delta_{H}^{2/3}$ with a large variance.  Is it possible that
accounting for this relationship  would be able to resolve the
discrepancy in $\xi_{\eCIV}(v)$?
In order to address this question, we repeated our experiment,
assuming that the local density followed the best-fit relationship
derived by Schaye \etal (2003).  Again we considered both high and low
metallicity models, resulting in correlation functions and column
density distributions that are shown in Fig.\ 13.

As our results, which depend on a line-fitting procedure, are biased
to the densest regions, the ``zero point'' metallicity of these models
are naturally shifted to lower values.  Thus,  the $\Delta^{2/3}
10^{-2} Z_\odot$ and the $\Delta^{2/3} 10^{-3} Z_\odot$ models shown
in these figures yield similar numbers of components as the single
metallicity $10^{-1} Z_\odot$ and $10^{-2} Z_\odot$ models,
respectively.  In particular, the lower metallicity model allows us to
obtain  good agreement with the observed column density distribution,
while assuming mean metallicity values more in line with previous
estimates (\eg Hellsten \etal 1997; Rauch, Haehenelt, \& Steinmetz
1997; Schaye \etal 2003).

Introducing a $\Delta$ dependence has almost no  effect, however, on
the correlation function, neither boosting it at low separations nor
introducing a feature at $\approx$ 150 km s$^{-1}$.   Thus is appears that this
nonlinear relationship is not the source of the clustering properties
of the metal-line components, and rather that the large variance seen in
the pixel-by-pixel results hides a third parameter  that determines
these features.  In fact, in Paper I, we described just such a key
parameter, the separation from a large dark-matter halo.

\begin{figure*}
\begin{center}
\resizebox{11.0cm}{!}{\includegraphics{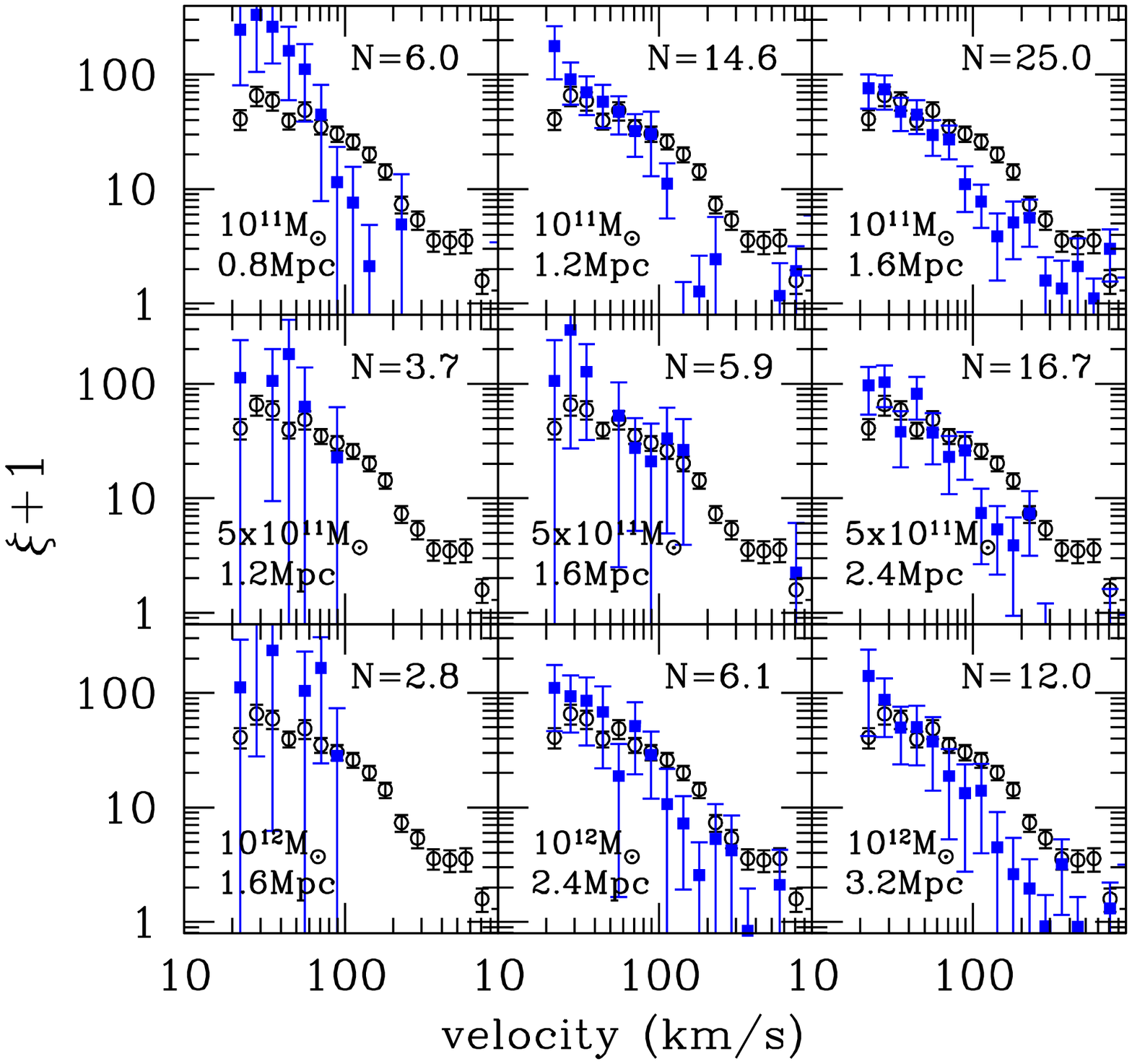}}\\%
%\centerline{\psfig{figure=f15.eps,height=10.0cm}}
\caption{Correlation function of simulated \CIV components,  with an
assumed bubble metallicity of $1/20 Z_\odot$.  Panels are labeled by
their assumed $M_s$ in units of $\msun$  and $R_s$ in units of
comoving Mpc.  In each panel the open circles summarize the
observational results,  while the filled circles represent the
experimental results, as averaged over 114 spectra.  Each panel is
labeled by the average number of \CIV components detected per simulated
spectrum.}
\end{center}
\label{fig:simcorr}
\end{figure*}

\begin{figure*}
\begin{center}
\resizebox{11.0cm}{!}{\includegraphics{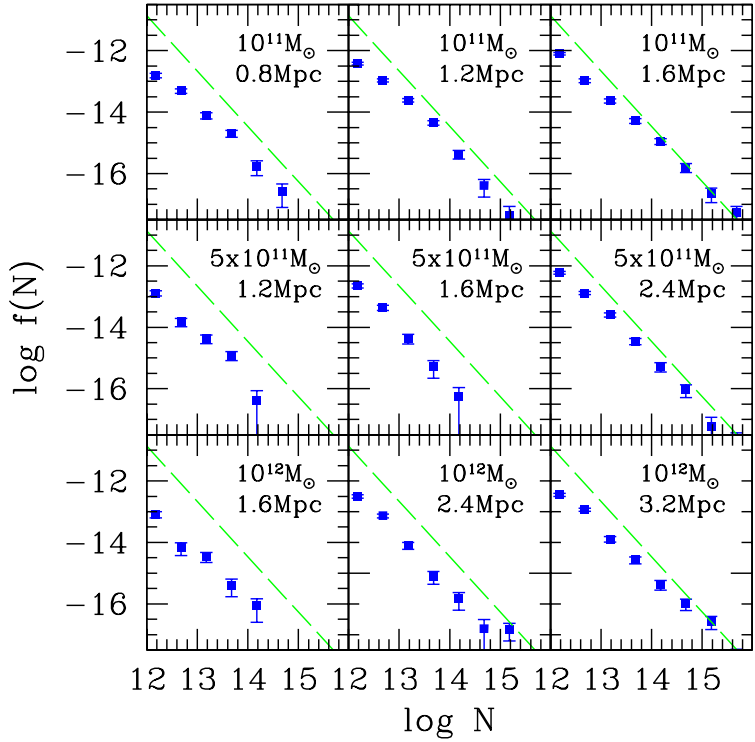}}\\%
%\centerline{\psfig{figure=f16.eps,height=10.0cm}}
\caption{Column density distribution of simulated \CIV absorption
components, with an assumed bubble metallicity of $1/20 Z_\odot$.
Models are as in Fig.\ 16 and in each panel the filled
points represent the simulation results, while the dashed line is the
fit given in Figure \ref{fig:abundance}.}
\label{fig:simcounts}
\end{center}
\end{figure*}

\begin{figure*}
\begin{center}
\resizebox{11.0cm}{!}{\includegraphics{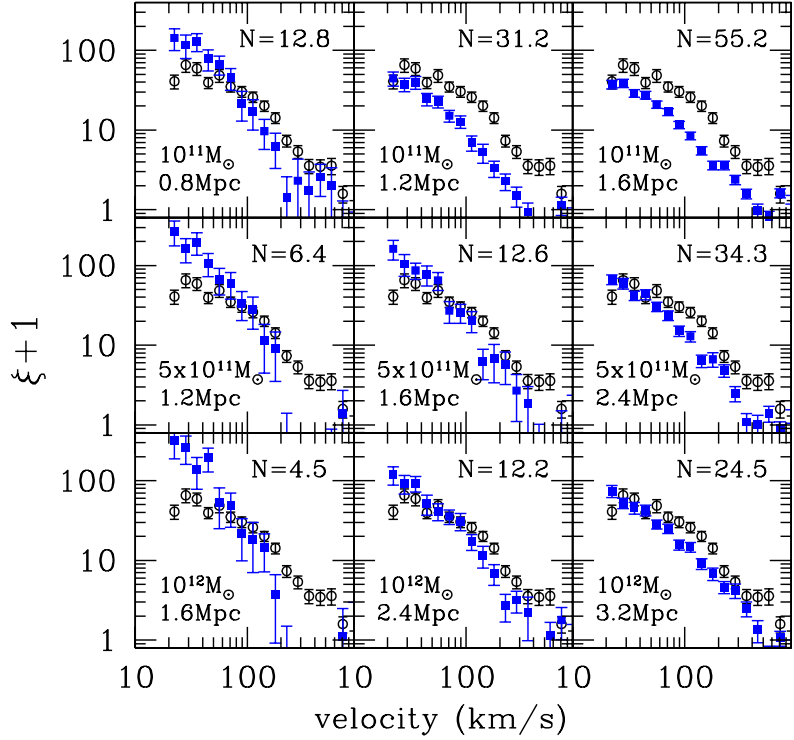}}\\%
%\centerline{\psfig{figure=f17.eps,height=10.0cm}}
\caption{Correlation function of simulated \CIV components,  with an
assumed bubble metallicity of $1/5 Z_\odot$.  Panels and symbols are
as in Fig.\ 16.}
\label{fig:simcorr2}
\end{center}
\end{figure*}

\begin{figure*}
\begin{center}
\resizebox{11.0cm}{!}{\includegraphics{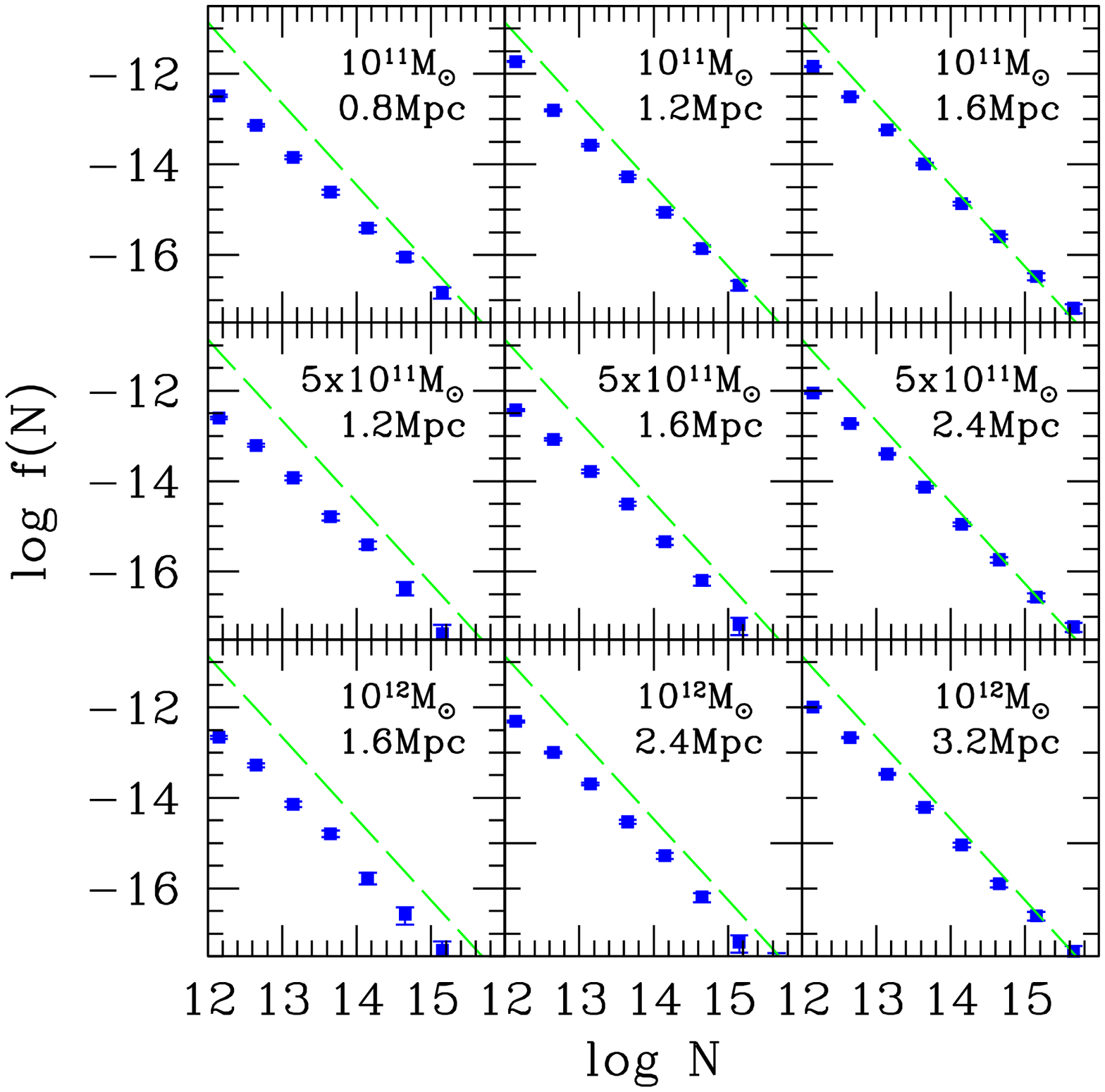}}\\%
%\centerline{\psfig{figure=f18.eps,height=10.0cm}}
\caption{Column density distribution of simulated \CIV absorption
components, with an assumed bubble metallicity of $1/5 Z_\odot$.
Panels and symbols are as in Fig. 17.}
\label{fig:simcounts2}
\end{center}
\end{figure*}

\section{Sources of Intergalactic Carbon IV}
 
\subsection{Distribution of Metals and Identification of Sources}

While the observed features in the \CIV correlation function can not
be understood in terms of a local nonlinear  relationship between the
metal and density distributions,  we saw in Paper I that these
features could be easily  explained in terms of the  distribution of
metal {\em sources}.  
In that work we used a pure dark-matter model to
describe \CIV components as contained within bubbles
centered around sources, and we interpreted  the 
amplitude and the knee in the \CIV correlation function in
terms of the source mass and bubble size, respectively.  
In this investigation we develop a similar  model, but
make use of the full gas and CLOUDY modeling  described in \S5 and \S6.

Following Paper I, we adopt a parameterization in which all metals are
found within a comoving radius $R_s$ of a dark matter halo whose mass is
above a fixed value, $M_s$.  To facilitate comparison
with our previous modeling, as well as to allow for future comparisons
with analytic approaches, we identified all sources 
at a fixed redshift of $z=3$.  In particular, halo
detection was performed using the HOP algorithm (Eisenstein \& Hut
1998) with parameters $\delta_{\rm peak}=160$, $\delta_{\rm
saddle}=140$, and $\delta_{\rm outer}=80$.  The centres of these
groups were then traced forward in time to the $z= 2.5$ and 2.0 slices
such that exactly the same groups could be selected  from all the
simulation slices, accounting for appropriate peculiar  motions.

As in Paper I, our choice of $z=3$ is not meant to imply that
enrichment occurred at this redshift, but rather that it occurred at
an unknown redshift higher than the observed range, centered on groups
whose large-scale clustering was equivalent to $z=3$ objects of mass
$M_s$.  For each choice of $R_s$ and $M_s,$ we then painted
bubbles of metals on our simulations, as illustrated in Fig.\ 14.
While increasing $R_s$ has the obvious effect of
increasing the volume filling factor of metals, increasing $M_s$ not
only lowers this filling factor, but also clusters the bubbles more
strongly.  This can be most easily seen by comparing models with
similar filling factors.  For example,  comparing the $M_s = 1 \times
10^{11} \msun,$  $R_s = 1.6$ comoving  Mpc slice to the  $M_s = 5
\times 10^{11} \msun,$  $R_s = 2.4$ comoving Mpc slice indicates that
a similar fraction is enriched with metals in both cases, but these
regions are  spread over a considerable area in the lower-mass case
and concentrated into dense knots in the higher-mass case.

\subsection{Properties of Carbon-IV}

From slices such as those shown in Fig.\ 14 we were able
to generate simulated QSO absorption spectra, in a manner  exactly
parallel to that described in \S 6.2: drawing lines-of-sight for the
various time outputs, piecing them together by evolving the mean
density, interpolating between CLOUDY tables, and applying realistic
levels of Poisson noise.  In this case the metallicity was assumed to
be at a fixed value $Z_b$ within each bubble, and zero everywhere
else.  Note, however, that our measurements are insensitive to 
\CIV components with columns below $10^{12}$ cm$^{-2}$, 
a thus a more widely dispersed, lower-level
contribution to IGM metals (\eg Schaye \etal 2003; 
Bergeron \& Herbert-Fort 2005) can not be excluded.

In Paper I, our modeling made use of a parameter $\tilde b$
that controlled the impact parameter associated with each sub-clump
within a bubble.  In our more physical modeling, this role is played
by $Z_b$, which we fixed at an initial value of $1/20 \zsun.$ These
spectra were again analyzed by our automated procedure, and in Figures
15 and 16 we compare the resulting
correlation functions and column density distributions with those
measured in the Large-Programme data set.

These plots reflect the trends seen in the slices.  Increasing the
mass concentrates the metal into fewer regions, boosting the
correlation function, particularly at large separations.  Increasing
$R_s,$ on the other hand, impacts the correlation function primarily
at smaller separations, and has a strong impact on the total number 
of \CIV components detected per spectrum.  From Fig.\ 15,  
the best-fit  models are the $M_s = 5 \times
10^{11} \msun$ \& $R_s = 2.4$ Mpc, $M_s = 10^{12} \msun$ \&  $R_s = 2.4$ Mpc,
and  $M_s = 10^{12} \msun$ \&  $R_s = 3.2$ Mpc cases, with filling factors
of 5.5\%, 8.6\%, and 11.6\% respectively although several of the
lower filling-factor cases produce so few lines as to be difficult to
evaluate in detail.  Similar filling factors have been advocated
by Pieri \& Haehnelt (2005) on the basis of O{\sc ~vi} measurements.
The large $M_s$ values we derive 
are also suggestive of the regions around Lyman-Break galaxies (LBGs),  
which are observed
to be clustered like $\approx 10^{11.5} \msun$ halos at $z = 3$
(Porciani \& Giavalisco 2002), and for which a strong cross-correlation
with \CIV absorbers has been measured (Adelberger \etal 2003).
It is also reminiscent of the association between galaxies and \CIV
absorbers put forward in BSR03.

In Fig.\ 16  we better quantify the number of components
in each model, by constructing simulated column density distributions
as discussed in  \S3.1.  Here we see that regardless of our choice of
source mass and bubble radius, all these models fall short of
reproducing the observations.   Due to the
relatively small filling factors of such bubble models,  our choice
of $Z_b = 1/20 Z_\odot$ is not able to generate the  relatively
large number of \CIV absorbers seen in the data.

In order to improve this agreement then, we considered a model in
which we assume a higher bubble metallicity of $Z_b \approx 0.2 Z_\odot$,
generating the $\xi_{\eCIV}(v)$ and $\log f(N)$ values seen in Figures
17 and 18.  As was the case for our
$\tilde b$ parameter in Paper I, varying $Z_b$ has relatively little
impact on the \CIV correlation function, although the increased number of
components does result in less noise.

Thus the high-metallicity simulations display the same trends  and
best-fit models as were seen in the lower $Z_b$ case, however the
improved signal allows us to distinguish the $M_s = 10^{12} \msun,$ $R
= 2.4$ Mpc model as a somewhat better match to the data than the  $M_s
= 5 \times 10^{11} \msun,$ $R = 1.6$ Mpc and $M_s = 10^{12} \msun,$ $R
= 3.2$ Mpc  models. The improved signal in  Fig.\ 17
also enables us to reject cases with very low filling factors. In
particular, we see that  the  models with the smallest bubble sizes do
not reproduce the observed $\approx 150$ km s$^{-1}$ elbow, exceeding the
measured $\xi_{\eCIV}(v)$  at small separations.  Furthermore, models with
$M_s = 10^{11} \msun$  are now seen to fall far short of the
observed  correlation function at large separations, particularly if
we consider the models with $R_s \geq 0.8$, which are not overly
peaked at small distances.

Finally, assuming a mean bubble metallicity of $1/5 Z_\odot$ has a
large  impact on the column density distribution, approximately
doubling the number of detected components and bringing our best-fit
model in rough agreement  with the data, although perhaps even this
value is slightly low in our  best-fit cases.  It is clear that we are
forced toward these values because much of the gas around $\approx
10^{12} M_\odot$ is heated by infall above $\approx 10^{5.5}$ K, and thus it
is largely in the outskirts of our bubbles in which \CIV absorbers are
found.

While, at face value, this metallicity is widely discrepant with other 
estimates, there are nevertheless two reasons to take it
seriously.   Firstly, the dense metal-rich regions in our model are
observed to be enriched to similar levels at $z=1.2$.  At this point,
the LBG-scale halos around which we have placed our metals are
expected to have fallen into clusters, and thus the IGM gas is
detectable through X-ray emission in  the intracluster medium (ICM).
In fact, detailed Chandra and XMM-Newton observations indicate ICM
iron levels of $Z = 0.20^{+0.10}_{-0.05}$ at $z=1.2$ (Tozzi \etal
2003), implying that at even higher redshifts these metals have been
efficiently  mixed into the diffuse gas that forms into clusters.
Secondly, we note that more than enough star formation has occurred by
$z=2.3$ to enrich these regions to our assumed values.  Indeed,
comparisons between the integrated star formation history and more
standard estimates of IGM metallicity have shown that a large fraction
of $z \approx 2$ metals have so-far escaped  detection (Pettini 1999).
Thus, we find no compelling reason to dismiss this high metallicity
value as spurious, although we emphasize that it has no direct impact
on our derived clustering masses and bubble sizes.

\section{An Analytic Model}

While our simulated bubble model provides a compelling picture of the
\CIV and \SiIV distribution at $z \approx 2-3$, it leaves open the
question as to properties of \MgII and {Fe{\sc ~ii}}. Yet the detailed
modeling of these species is beyond the capabilities of our
simulation.   As we saw in \S6.1, the environments of \MgII and \FeII
are denser than  \CIV and {Si{\sc ~iv}}, particularly  in the case of
{Fe{\sc ~ii}}.  Even more constraining is the fact that almost all our
detections of these systems fall well below our final simulated
redshift of 2, with the majority lying in the $0.5 \lesssim z \lesssim 1.5$
redshift range.

In Fig.\ \ref{fig:corrall} we saw that while the overall shape and
correlation function of these species is comparable to that observed
in \CIV and {Si{\sc ~iv}}, the magnitude and long-separation 
tail of $\xi(v)$ are shifted upwards by a factor associated with the 
cosmological growth of structure.  While reproducing 
these trends is beyond the capabilities of our simulation,
they can nevertheless be studied from an approximate
analytic perspective.

\subsection{Derivation}

In \S7 we found good agreement between our observations and a model in
which we painted metals around biased regions in our simulation.
Analytically this corresponds to a picture in which  the metal lines
observed at $z_{\rm obs}$ come from clumps that are within a
fixed radius of the sources of IGM metals. These pollution sources are
associated with relatively rare objects of mass $M_p$ that ejected
metals  into  surroundings at a high redshift $z_{p} > z_{\rm
obs}$.  After enrichment these components continue to cluster
gravitationally to $z_{\rm obs}$.

In the numerical simulations, these pollution centres are
identified with the galaxies of mass $M_s$  at a redshift of $z=3$.
However, this mass and redshift were intended only to quantify the
bias of sources, and it is more probable that they
are really related to less massive, higher redshift objects, which
exhibit similar  clustering properties (Scannapieco 2005).

Let us consider then four points: the centres of two  clumps (1 and
2), which we observe as metal-line components, and the centres of two
bubbles  (3 and 4), which correspond to the sources of pollution. We
require  that the pollution sources correspond to peaks [\ie linear
overdensities  with a contrast larger than 
$\delta_{\rm cr} \equiv 1.68 D(z_{p})^{-1}$]
at a redshift  $z_{p} > z_{\rm obs}$ and at a mass scale 
$M_{p}$.  The clumps, on the other hand, are associated with the \CIV
absorbers themselves, and correspond to peaks at a mass scale $M_c$.  In
the linear approximation, these fields satisfy a joint Gaussian
probability distribution which is specified by the  block correlation
matrix:
\ba
\M{M}& = & \left[   \matrix{  {{\xi  }_{\Mvariable{cc}}}(0)   &  
\,{{\xi       }_{\Mvariable{cc}}}(r_{12})      &             \,{{\xi
}_{\Mvariable{cp}}}(r_{13})  &  \,{{\xi  }_{\Mvariable{cp}}}(r_{14}) \cr
 \,{{\xi  }_{\Mvariable{cc}}}(r_{12}) & {{\xi }_{\Mvariable{cc}}}(0)
&      \,{{\xi   }_{\Mvariable{cp}}}(r_{23})  &      \,{{\xi
}_{\Mvariable{cp}}}(r_{24}) \cr   \, {{\xi }_{\Mvariable{cp}}}(r_{13}) &
 \,{{\xi  }_{\Mvariable{cp}}}(r_{23}) & {{\xi }_{\Mvariable{pp}}}(0)
&      \,  {{\xi   }_{\Mvariable{pp}}}(r_{34})   \cr     \,{{\xi
}_{\Mvariable{cp}}}(r_{14})   &     \,{{\xi   }_{\Mvariable{cp}}}(r_{24})  &
 \,{{\xi }_{\Mvariable{pp}}}(r_{34})  & {{\xi }_{\Mvariable{pp}}}(0) \cr
} \right] \nonumber \\
& \equiv &  \left[ \matrix{ \M{M}_\Mvariable{cc} & \M{c}_\Mvariable{cp} \cr
\M{c}_\Mvariable{pc} & \M{M}_\Mvariable{pp} } \right] \,,
\ea
where $r_{ij}\equiv || \M{r}_{i}-\M{r}_{j}|| $, and $\xi_{\rm pp}$, 
$\xi_{\rm cc}$ and $\xi_{\rm cp}$
refer respectively to the correlation between pollution centres,
satellite clumps, and the cross 
correlation between clumps and the centres.  
The joint probability of having a peak of an amplitude larger than 
$\delta_{\rm cr}$ at the four points is given by
\ba
p(1,2,3,4) &=&
\frac{1}{4 \pi^{2} \sqrt{\det|\M{M}|}}   
\int_{\delta_{\rm cr}}^{\infty} \d \delta_{1} 
     \int_{\delta_{\rm cr}}^{\infty} \d \delta_{2}  \\
 & \times&    \int_{\delta_{\rm cr}}^{\infty} \d \delta_{3} 
        \int_{\delta_{\rm cr}}^{\infty} \d \delta_{4}\nonumber \\ 
  & \times&   e^{-\frac{\left(
    \delta_{1},   \delta_{2},   \delta_{3},   \delta_{4}
    \right)^{T}\cdot \M{M}^{-1} \cdot \left(
    \delta_{1},   \delta_{2},   \delta_{3},   \delta_{4}
    \right)}{2}}. \qquad \nonumber
    \label{eq:defp}
\ea
We will evaluate this expression, assuming that the threshold that defines
the object is high relative to the corresponding {\it rms} densities and
taking the correlation between  the metal line clumps and 
the centres of pollution to be small. 
We shall not assume the smallness of the centre-centre 
nor clump-clump correlation, of which the first is the most important.
In this limit 
\ba
\M{M}^{-1} & \approx & \left[ \matrix{ \M{M}^{-1}_\Mvariable{cc} &  -
\M{M}^{-1}_\Mvariable{cc} \M{c}_\Mvariable{cp}
\M{M}^{-1}_\Mvariable{pp} \cr - \M{M}^{-1}_\Mvariable{pp}
\M{c}_\Mvariable{pc}  \M{M}^{-1}_\Mvariable{cc} &
\M{M}^{-1}_\Mvariable{pp} } \right] \,, \\
 \det \left| \M{M} \right| 
& \approx &  \det \left| \M{M}_\Mvariable{cc} \right| \cdot \det \left|
\M{M}_\Mvariable{pp} \right|,
\ea
and
\begin{eqnarray}
p(1,2,3,4) &\approx & \frac{1}{4 \pi^{2} 
\sqrt{\det|\M{M_{cc}}|} 
\sqrt{\det|\M{M_{pp}}|}} \nonumber \\
 &\times& \int_{\delta_{\rm cr}}^{\infty} \d \delta_{1}
\int_{\delta_{\rm cr}}^{\infty} \d \delta_{2}
e^{-\frac{1}{2}\left(\delta_{1},\delta_{2}\right)^{T}
\cdot\M{M}^{-1}_\Mvariable{cc} \cdot
\left(\delta_{1},\delta_{2}\right)} \nonumber \\ 
&\times&
\int_{\delta_{\rm cr}}^{\infty} \d \delta_{3}  \int_{\delta_{\rm cr}}^{\infty} \d
\delta_{4}
e^{-\frac{1}{2}\left(\delta_{3},\delta_{4}\right)^{T}
\cdot\M{M}^{-1}_\Mvariable{pp} \cdot
\left(\delta_{3},\delta_{4} \right)} \nonumber \\ & \times & 
e^{\left(\delta_{1},\delta_{2}\right)^{T} \cdot
\M{M}^{-1}_\Mvariable{cc} \M{c}_\Mvariable{cp}
\M{M}^{-1}_\Mvariable{pp} \cdot
\left(\delta_{3},\delta_{4}\right)}. 
\label{eq:defapprox}
\end{eqnarray}
In the high peak limit, the last cross-correlation term can be
factored out from the integrals (see Appendix) yielding
\ba
p(1,2,3,4) \approx  p(1,2) p(3,4)
e^{\left(\delta_{cr},\delta_{cr}\right)^{T} \cdot
\M{M}^{-1}_\Mvariable{cc} \M{c}_\Mvariable{cp}
\M{M}^{-1}_\Mvariable{pp} \cdot
\left(\delta_{cr},\delta_{cr}\right)}, 
\ea
where $p(1,2)$ and $p(3,4)$ are computed from eq.\ (\ref{eq:p12uni}).
Or, explicitly,
\begin{eqnarray}
\label{eq:p1234}
p(1,2,3,4)&\approx &\frac{1}{4 \pi^2} \nu_\Mvariable{cc}^{-2}
\nu_\Mvariable{pp}^{-2}
\Mvariable{C}\left[c(r_{12}),\nu_\Mvariable{cc}\right]
\Mvariable{C}\left[c(r_{34}),\nu_\Mvariable{pp}\right] 
\\ &\times& \exp\left[
-\frac{\nu_\Mvariable{cc}^2}{1+c_\Mvariable{cc}(r_{12})}
-\frac{\nu_\Mvariable{pp}^2}{1+c_\Mvariable{pp}(r_{34})}  \nonumber
+ \nu_\Mvariable{cc}\nu_\Mvariable{pp} \right. \nonumber \\
& \times & \left.
\frac{c_{\Mvariable{cp}}(r_{13})+ c_{\Mvariable{cp}}(r_{24})+
c_{\Mvariable{cp}}(r_{14})+c_{\Mvariable{cp}}(r_{23})}
{\left[1+c_{\Mvariable{pp}}(r_{34})\right]
\left[1+c_{\Mvariable{cc}}(r_{12})\right]} \right], \nonumber 
\end{eqnarray}
where the function $\Mvariable{C}(x,\nu)$ is defined in the Appendix,
and  we define the cross-correlation coefficients as
$c_\Mvariable{cc}\equiv\xi_\Mvariable{cc}(r)/\xi_\Mvariable{cc}(0)$,
$c_\Mvariable{pp}\equiv\xi_\Mvariable{pp}(r)/\xi_\Mvariable{pp}(0)$,
$c_\Mvariable{cp}\equiv\xi_\Mvariable{cp}(r)/\sqrt{\xi_\Mvariable{cc}(0)
\xi_\Mvariable{pp}(0)}$, and the normalized density thresholds as
$\nu_\Mvariable{cc}\equiv\delta_{\rm cr}/\sqrt{\xi_\Mvariable{cc}(0)}$ and
$\nu_\Mvariable{pp}\equiv\delta_{\rm cr}/\sqrt{\xi_\Mvariable{pp}(0)}$. 
\footnote{The cross-correlation coefficients
$\xi_{cc}(r)/\xi_{cc}(0)$ and $\xi_{pp}(r)/\xi_{pp}(0)$ reach unity at
$r=0$ and thus cannot be assumed small everywhere. At the same time
$\xi_{cp}(r)/\sqrt{\xi_{cc}(0) \xi_{pp}(0)}$ is always less than unity
if $M_c$ and $M_p$ do not coincide (Schwartz Inequality).  
In particular the smaller its maximum value, achieved at $r=0$, the larger
the difference between the scales describing 
the clumps and the pollution centres.}  

Note that in eq.\ (\ref{eq:p1234}) the correlation functions in  the
denominator are not assumed to be small, which allows for proper
accounting of the case when two clumps or two pollution centres
are the same. For
example, setting  $r_{12}=r_{34}=0$ properly recovers the bivariant
joint probability $p(1,3)$ to find a clump at a separation
$r_{13}$ from the centre of pollution
(equal in this case to $r_{14}=r_{23}=r_{24}$).

In our model only those clumps that lie within the spherical bubble around 
some pollution centre are observed to have metals.
The correlation function of clumps of mass $M_c$ that are
within spherical bubbles around peaks corresponding to the 
mass $M_p$, is defined as
\begin{equation}
p(\delta_1,\delta_3)p(\delta_2,\delta_4)
\left[1+\bar\xi(r_{12})\right]
\equiv \bar p(\delta_1,\delta_2,\delta_3,\delta_4),
\label{eq:corrdef}
\end{equation}
where the bar denotes averaging over position of pollution centres
within a distance $R_s$ around two metal-rich clumps
at a fixed separation $r_{12}$.
Note that our definition of the correlation function, $\bar \xi(r_{12}),$ 
is not equivalent to the estimator of the underlying correlation function 
of all the clumps of mass $M_c$, 
$\xi(r_{12})$, nor is it equivalent to the conditional correlation function 
if there were a source of metals (a high peak of the scale $M_p$)
in the vicinity of {\em every} small halo,
$\bar \xi_c(r_{12})$,  which would be given by
\begin{equation}
p(\delta_1)p(\delta_2)\bar p(\delta_3,\delta_4)
\left[1+\bar\xi_c(r_{12})\right]
\equiv \bar p(\delta_1,\delta_2,\delta_3,\delta_4).
\end{equation}
Furthermore, $\bar \xi(r_{12})$ 
depends on the underlying two-point
correlation of small clumps, the correlation of the sources, and the
cross-correlation between clumps and sources. This last term is 
subject to the most modification should the physics  of metal
dispersal change.  However it mostly affects the biased density
of small clumps in the vicinity of the sources relative
to the cosmological mean, which is precisely the excess
factored out in eq.\ (\ref{eq:corrdef}).

Thus eq.\ (\ref{eq:corrdef}) describes the correlation of 
metal components at the redshift of pollution, which
is dominated by the clustering of the pollution sources. Subsequent
gravitational clustering of enriched metals then leads to further
amplification of the  correlation in the linear approximation as
\be
\bar \xi(r, z_{\rm obs}) = 
\left[D(z_{\rm obs})/D(z_{p})\right]^2 \bar\xi(r,z_{p}),
\ee
where $D(z)$ is the linear density growth factor.  This growth is
suggestive of the difference between the \CIV and \MgII 
correlation functions, as we saw in Fig \ref{fig:corrall}, as well
as the hints of evolution seen in $\xi_{\eCIV}(v)$ 
and $\xi_{\eMgII}(v)$ in Figs. \ref{fig:CIV} and \ref{fig:MgIIcomp}.

\subsection{Application to observed metal absorbers}

\begin{figure}
\centerline{\psfig{figure=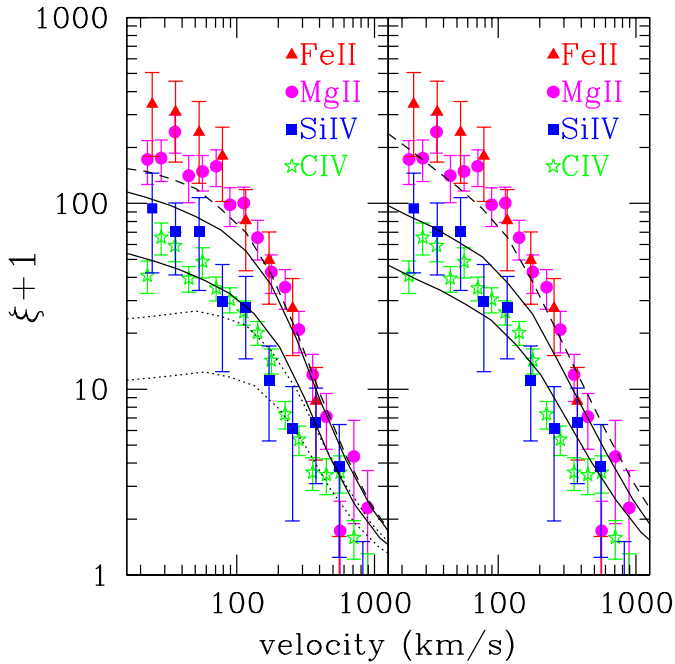,height=9.5cm}}

\caption{Comparison of our analytic model with the data.  {\em Left
panel:} Low-redshift model for metal sources.  Dotted lines represent
the linear clustering at  $z_{\rm obs} = 2.3$ (lower) and 
$z_{\rm obs} = 1.15$
(upper) of clumps  observed in the vicinity of the pollution centres,
with  $M_p= 10^{12} M_\odot$, $M_c=10^9 M_{\odot}$  and
$z_{p} =3$.  Solid lines show the effect of applying a non-linear
correction to these models.  Finally, the dashed curve shows a
nonlinear  $z_{\rm obs} = 1.15$ model in which $M_p= 10^{12}
M_\odot$, and $z_{p} =3,$ but now $M_c=10^{10} M_{\odot}.$  {\em
Right panel:}  Lower solid curve corresponds to  $M_p=3\times10^9
M_{\odot}$, $M_c=10^8 M_{\odot}$, $z_{\rm \rm pol}=7.5$, $z_{\rm
obs}=2.3$, with no  nonlinear correction applied.  The upper solid
line is  a further linear extrapolation of this model to $z_{\rm
obs}=1.15$.  The upper dashed line is a linear model again with
$z_{\rm obs}=1.15$ $M_p=3\times10^9 M_{\odot}$, $z_{\rm \rm pol}=7.5$,
but with $M_c$ raised to $10^9 M_{\odot}$.  For all curves the comoving size 
of the bubble is $2.4$~Mpc.}
\label{fig:analytics}
\end{figure}

In Figure \ref{fig:analytics}   we fit our analytic model to
the data. In the left panel we adopt the parameters used in our
numerical simulations, identifying metal pollution centres with $M_p =
10^{12} M_{\odot}$ objects at a redshift $z_{p} = 3$ and metal
rich clumps with collapsed objects of $M_c = 10^9 M_{\odot}$, with 
$R_s =2.4$ comoving Mpc.
At $z_{\rm obs} = 2.3,$ the analytic
fit reproduces the measured $\xi_{\eCIV}(v)$ at large velocity
separations, where it is dominated by the correlation between 
pollution centres, but it falls short at small velocities, where
$\xi_{\eCIV}(v)$ is dominated by the clump distribution within each
bubble.

This is because the smoothing imposed by choosing  $M_p \approx 10^{12}
M_{\odot}$ is similar to the 2.4 Mpc bubble size, and thus our
linear formalism is insufficient to describe distances less than $R_s$.
In reality, the nonlinear collapse of $M_p$ would have moved in
new material to fill in this region.  To mimic such nonlinear effects
at small radii, we apply the prescription $\delta_{\rm cr} \to
\delta_{\rm cr}+(1-1/\delta_{\rm cr}) \xi$ (Mo \& White 1996), resulting in the
dashed curve.   This correction, while crude, is seen to recover a
$\xi_{\eCIV}(v)$ that is similar to our simulated $10^{12}
M_\odot$, $R_s = 2.4$ Mpc case (and thus the observed correlation
function), confirming that  the discrepancy at small distances is
caused by our neglect of  nonlinear motions.

Next we turn our attention to  \MgII and {Fe{\sc ~ii}},
which  are observed at
lower redshifts $z\approx 1.2$.  As we saw in Figure
\ref{fig:corrall} the rise of the correlation amplitude of these
species relative to \CIV and \SiIV is generally in agreement with the
hypothesis of linear growth of
gravitational clustering of a fixed population of
objects from $z=2.3$ to $z=1.15$, although there are significant
discrepancies at small radii.
Again we plot both a linear $z_{\rm \rm pol} = 3$,  $M_p=
10^{12} M_\odot$, $M_c=10^9 M_{\odot}$  model observed at $z_{\rm
obs} = 1.15$  and a similar model in which a nonlinear correction has  been
applied.  While the nonlinear curve does well at most radii, a
shortfall is seen at $z \lesssim 100$ km s$^{-1}$, similar to the discrepancy between
the ``shifted''  $\xi_{\eCIV}(v)$ curve and the $\xi_{\eMgII}(v)$
curves in  Figure \ref{fig:corrall}.  Based on our plots of the
species fraction as a function of environment, an important difference
between these species is clear.  As \MgII can only survive in regions
with a low ionization parameter, it is biased toward much denser
regions than {C{\sc ~iv}}, which correspond in our analytic models to
higher clump masses.   Raising $M_c$ to $10^{10} M_{\odot}$ to
account for this effect leads to  the dashed curve in the left
panel, which again agrees well with the data.

As discussed above, however, it is likely that the origin of
metal pollution lies at higher redshift from sources of a lower
mass, whose comoving clustering properties are identical to  $M \approx
10^{12} M_{\odot}$ galaxies identified at $z=3$.  Indeed, these biased
high-redshift sources may be the progenitors
that later grew into large $z=3$ galaxies.
In the right panel of Figure \ref{fig:analytics} we explore such a
high redshift model, in which we take $z_{\rm \rm pol}=7.5$ and
$M_p=3\times 10^9 M_{\odot}$, so that the bias of our sources is
the same as for $M_p= 10^{12} M_{\odot}$, objects forming at
$z_{p} = 3.$   Adopting a similar $\nu_c$ as in the
$z_p=3$ case results in the solid curves.    As the smoothing due to the
Lagrangian radius associated with  $M_p = 3 \times 10^9$ is minimal, 
no nonlinear correction is necessary and our simple model provides a 
reasonable fit to the \CIV and \SiIV components observed  at $z=2.3.$

Similarly, extrapolation of the same objects to $z=1.15$, the
mean redshift for \MgII  and \FeII systems, matches their large scale
($v > 300$ km s$^{-1}$) correlations quite well, although the data at
these separations is sparse.  At small velocities the
difference between the environments of the two species becomes
important, and  linear scaling does not completely explain the
enhancement of correlation amplitude in \FeII and \MgII relative to
\CIV and Si{\sc ~iv}.  As in the low redshift case, if we  associate
these species with larger clumps,  the fit is much improved at small
radii, resulting in  the dashed curve.

In summary, our simple analytic model generally reproduces the
features seen in our simulations of the \CIV and \SiIV
correlation functions, although a nonlinear correction is necessary in
the $z_{p} = 3$ model.  Linearly extrapolating these
models to lower redshift results in a good fit to \MgII and \FeII at
large distances, although a fit at smaller distances requires us to use
larger clump masses, associated with denser environments.  
Finally, we find that there is a strong degeneracy between 
$M_p$ and $z_{\rm \rm pol},$  with a  family of sources with similar biases 
producing acceptable fits.

\section{Conclusions}

While intergalactic metals are ubiquitous, the details of how these
elements made their way into the most tenuous regions of
space remains unknown.  In this study we have used a uniquely large,
homogeneous, and high signal-to noise sample of QSO sightlines to pin
down the spatial distribution of  these metals  and combined this with
advanced automated detection techniques and a high-resolution SPH
simulation to pin down just what we can learn from this distribution.
Our study has been focused on four key species: \CIV and {Si{\sc
~iv}},  which serve as tracers of somewhat-overdense regions from
redshifts 1.5 to 3.0, and \MgII and  {Fe{\sc ~ii}}, which trace dense,
lower-redshift ($z=0.5-2.0$) environments.  No evolution in the column
density  distributions of any of these  species is detected.

In the high-redshift case, \CIV and \SiIV trace each
other closely. For both species, $\xi(v),$  exhibits a steep decline
at large separations, which is roughly consistent with the slope of
the  $\Lambda$CDM matter correlation function and the spatial
clustering of $z \approx 3$ Lyman break galaxies.  At separations below
$\approx 150$ km s$^{-1}$, this function flattens out considerably,
reaching a value of $\xi(v) \approx 50$, if $v \lesssim 50 $ km
s$^{-1}$.  Our data also suggests that  
$\xi_\eCIV(v)$ evolves weakly with redshift, at a level
consistent with the linear growth of structure.

The distribution of metals as traced by
$\xi_\eCIV(v)$ is extremely robust.  We find that it remains almost
completely unchanged when minimum or maximum column density cuts are
applied to our sample, even if they are so extreme as to eliminate
over two-thirds of the components.  We have also linked
together \CIV components into systems, using a one-dimensional
friends-of-friends algorithm, with linking lengths of $v_{\rm link} = $
25, 50, and 100 km s$^{-1}$.  In all cases, the line-of-sight
correlation function of the resulting systems matches the original
component correlation function (within measurements errors) at
separations above  $v_{\rm link}.$ Finally, the {Si{\sc ~iv}}/\CIV
ratio shows no clear dependence when binned as a function of
separation, suggesting that the features seen in $\xi_\eCIV(v)$ and
$\xi_\eSiIV(v)$ do not result from fluctuations in the ionizing
background.

Thus none of our tests indicate that the observed distributions of
\CIV and \SiIV represent anything but the distribution of
intergalactic metals at $z = 1.5-3.0$.  This motivated us to carry out
a confrontation between our \CIV observations and detailed simulations
of IGM metal enrichment, which   paralleled previous comparisons for
the Lyman-alpha forest.  Furthermore, the advanced automatic-detection
procedures described in \S2.2 (see also Aracil \etal 2005) allowed us
not only to compare simulated and observed spectra, but generate
simulated  line lists in a manner that exactly paralleled the
observations.

Using these tools, we found that the observed features of the \CIV
line-of-sight correlation function can not be reproduced  if the IGM
metallicity is constant.   Rather any
such model falls far short of the observed $\xi_\eCIV(v)$ amplitude
and fails to reproduce flattening seen below $\approx 150$ km s$^{-1}.$
Furthermore, adopting a local relation between
overdensity and metallicity, as observed by Schaye \etal (2003), has
little or no effect on these  results.

On the other hand, rough agreement between simulated and observed
\CIV correlations is obtained in a model in which only a {\em
fraction} of the IGM is enriched.  Emulating the simple model in Paper
I, we explored a range of models  in which metals were confined within
bubbles of radius $R_s$ about $z=3$ sources of mass $M_s$, where these
quantities are not meant literally as source redshifts  and masses,
but rather as tracers of the {\em bias} of the $z_{\rm \rm pol} \geq
3$  source population.   Varying these quantities, we
derived parameters that suggest large metal bubbles, $R_s \approx 2$
comoving Mpc, around highly-biased sources, with $M_s \approx 10^{12}
\msun.$

These results are suggestive of the association between galaxies and \CIV 
absorbers put forward in BSR03, and 
the high cross-correlation between
LBGs and \CIV absorbers measured by Adelberger \etal (2003).
Yet this does not mean that LBGs are the sources of
intergalactic metals, only that the  $z_{\rm \rm pol} \geq 3$
sources were biased like LBGs.  In fact, the case for outflows
escaping  lower-redshift starbursts is much more convincing than for
dwarves (Martin 2005).  Similarly, $R_s$ need not be interpreted as
the ejection radius of each source, but instead as the distance at
which bubbles from multiple sources overlap.   Our best fit $M_s$ and
$R_s$ values are independent of the assumed bubble metallicity,
although the low $\lesssim 10\%$ volume filling factors of these
models forces us to use large $\approx 1/5 Z_\odot$ values to reproduce the
observed \CIV column density distribution.  Note however that given
the high bias of our enriched regions, such metallicities may be
necessary to reconcile $z \approx 2.3$ measurements with $z \approx 1.2$
observations of the iron content of the ICM in high-redshift galaxy
clusters (Tozzi \etal 2003).

At lower redshifts, the line-of-sight correlation functions  of \MgII
and \FeII are consistent with the same  enriched regions seen in
\CIV and {Si{\sc ~iv}},  but  now ``passively'' evolved down to $z
\approx 1.2.$ Again both $\xi_{\eMgII}(v)$ and $\xi_{\eFeII}(v)$ trace
each other closely, and exhibit the same steep decline at large
separations and flattening at small separations as were seen in
$\xi_{\eCIV}(v)$ and $\xi_{\eSiIV}(v).$   Also as in the higher
redshift case, the \MgII correlation function remains unchanged when
minimum and maximum column density cuts are applied, and linking
together \MgII components into systems has no strong impact on
$\xi_{\eMgII}(v)$ outside separations corresponding to the linking length.

Although \MgII and \FeII are detected in regions that can not be
simulated numerically, we are nevertheless able to develop  an
analytic model that allows for a simple analysis of these species.
Testing our model against $\xi_{\eCIV}(v)$ and $\xi_{\eSiIV}(v)$, we
find generally good agreement with the data for similar values of mass and
$R_s$ as in the numerical case.  Pushing the model to lower redshift,
we find that the  same parameters do well at reproducing the
clustering properties of \MgII and {Fe{\sc ~ii}}, especially when we
account for the fact that the species are found in denser
environments.  Finally, we also find agreement with the observed
$\xi_{\eCIV}(v)$ and $\xi_{\eSiIV}(v)$ at $z=2.3$ and
$\xi_{\eMgII}(v)$ and $\xi_{\eFeII}(v)$ at $z=1.15$ and a
high-redshift analytic model in which $z_{p} = 7.5$ and $M_p =
3 \times 10^9 \msun,$ illustrating the strong degeneracy between $M_p$
and $z_{p}$ for similarly biased sources.

Taken together, our $z \approx 2.3$ and $z \approx 1.2$ measurements,
numerical simulations, and analytic modeling paint a consistent
picture of IGM enrichment.  The distribution of intergalactic metals
does not appear uniform, nor simply dependent on the local density, but
rather it bears the signature of the population from which it came.
While the $z \geq 3$ redshift of metal ejection is unknown,  
a joint constraint on the masses and redshifts of the objects
responsible for IGM pollution remains compelling.  
Models of IGM enrichment must come to terms with the observed biased 
sources of intergalactic metals.

\section*{Acknowledgments}

We are grateful to K.~Adelberger, A.~Aguirre, D.~Aubert, M.~Davis,
A.~Ferrara, M. Haehnelt, J.~Heyvaerts, C.~Martin, C.~Mallouris,
E.~Rollinde, J.~Schaye, R.~Teyssier, and E.~Thi\'ebaut, for useful
comments and helpful suggestions.  We thank the anonymous
referee for a careful reading of our paper, which greatly improved it.
ES was supported in part by an NSF
MPS-DRF fellowship.  RJT\ acknowledges funding from the Canadian
Computational Cosmology Consortium and use of the SHARCNET  computing
facilities.  This work greatly benifited from the many  collaborative
discussions made possible by visits by ES and DP, hosted by the
Observatoire de  Strasbourg.  We would like to thank D.~Munro for
freely distributing his Yorick programming language (available at
{\em\tt http://www.maumae.net/yorick/doc/index.html}) which we used to
implement our algorithm, and F.~Haardt \& P.~Madau for providing us
with an updated version of their UV background models.  This work is
based on observations collected through ESO project ID No.
166.A-0106.  We acknowledge partial support from the Research and
Training Network `The Physics of the Intergalactic Medium' set up by
the European Community under the contract HPRN-CT2000-00126 RG29185.
This work was supported by the National Science Foundation under
grant PHY99-07949.

\appendix\onecolumn
\section{Two point joint probability distribution of high peaks
in Gaussian fields}

In this section we present formulas for the joint two point 
probability for the peaks in the Gaussian field
that are used in our analytic model. These allow for the pair of
peaks to have different scales and improve on the asymptotic results
for the high peaks.

In the standard cosmological picture one identifies a collapsed object
of mass $M$ with a peak of  height $\delta > \delta_{cr}$
in the density field $\delta(x)$, smoothed with a top-hat
window function $W(R)$ with the scale 
$R=\left(\frac{3 M}{4 \pi \bar \rho}\right)^{1/3}$.
In the limit of large height the geometrical
peaks of the Gaussian field can be approximately described as
just the regions of high field values. This is the approximation that we
adopt.

We shall need, first, the variance of the smoothed density field
\begin{equation}
\sigma^2 = \int P(k) W^2(k R)  k^{2} \d k,
\end{equation}
where $P(k)$ is the power spectrum of the density field and the
Fourier image of the top-hat window is
\be
W(k R) \equiv 4 \pi R^3 
\left[ \frac{\sin k R}{(K R)^3} - \frac{\cos k R}{(K R)^2} \right],
\ee
and, second, the correlation function
between the values of the field at two positions, 
separated by the distance $r_{12}=x_1-x_2$
\be
\xi(r_{12})=\int P(k) \frac{\sin(k r_{12})}{k r_{12} }
W(k R_{1}) W(k R_{2}) k^{2} 
\d k,
\ee
where the value at point $1$ is taken after the field is smoothed on a 
scale $R_1$ while at the point $2$ the field is evaluated after smoothing
on a scale $R_2$. If $R_1=R_2$, then $\xi(0)=\sigma^2$, while in general
$\xi(0) \le \sigma_1 \sigma_2$.

To evaluate the probability distribution functions used in \S8
we begin with the well-known result for the one point probability of
the field height to exceed $\delta_{cr}:$
\begin{equation}
p(1) = \frac{1}{\sqrt{2 \pi} \sigma} \int_{\delta_{\rm cr}}^{\infty} \d \delta_1
\exp\left[-\frac{\delta_1^2}{2 \sigma_1^2}\right] \sim
\frac{1}{\sqrt{2 \pi}} \frac{\sigma_1}{\delta_{\rm cr}}
\exp\left[-\frac{\delta_{\rm cr}^2}{2 \sigma_1^2}\right]
=\frac{1}{\sqrt{2 \pi}} \nu_1^{-1}
\exp\left[-\frac{\nu_1^2}{2}\right],
\label{eq:def1p}
\end{equation}
where $\nu_1 \equiv \delta_{\rm cr}/\sigma_1.$
Here $1$ refers both to the (arbitrary) point where the field is evaluated,
as well as to the scale it was smoothed with, $R_1$.

Next we evaluate the asymptotic behaviour at large 
$\delta_{\rm cr} \gg \sigma$ for the joint two-point probability
\begin{equation}
p(1,2)=
\frac{1}{2 \pi \sqrt{\sigma_1^2\sigma_2^2-\xi(r_{12})}}
\int_{\delta_{\rm cr}}^{\infty} \d \delta_{1} \int_{\delta_{\rm cr}}^{\infty} \d \delta_{2} 
\exp\left[-\frac{1}{2}
\frac{\delta_1\sigma_2^2 + \delta_2\sigma_1^2 - 2 \xi(r_{12})\delta_1\delta_2}
{\sigma_1^2\sigma_2^2-\xi^2(r_{12})}\right]~,
\label{eq:defp2}
\end{equation}
paying attention to the prefactors to the exponential terms.
In general $\sigma_1 \ne \sigma_2,$  but when $\delta_1$ and 
$\delta_2$ represent the same field smoothed with the same filter taken
at two different points (the case that we mostly need in this paper)
$\sigma_1=\sigma_2$. 
Introducing uncorrelated variables 
$x=\frac{\delta_1 \sigma_2 +\delta_2 \sigma_1}
{\delta_{\rm cr} \sigma_2 +\delta_{\rm cr} \sigma_1} $
and $y=\frac{\delta_1\sigma_2-\delta_2\sigma_1}{\sigma_1\sigma_2}$,
we obtain
\begin{equation}
p(1,2)=
  \frac{\delta_{\rm cr} (\sigma_1+\sigma_2)}
{4 \pi \sqrt{\sigma_1^2\sigma_2^2-\xi^2(r_{12})}}
\int_{1}^{\infty} \d x \;
\exp\left[-\frac{1}{4} \frac{\delta_{\rm cr}^2}{\sigma_1\sigma_2}
\frac{(\sigma_1+\sigma_2)^2}{\sigma_1\sigma_2+\xi(r_{12})} x^2\right]
\int_{\frac{\delta_{\rm cr}[2\sigma_2-x(\sigma_1+\sigma_2)]}{\sigma_1\sigma_2}}^{\frac{\delta_{\rm cr}[x(\sigma_1+\sigma_2)-2\sigma_1]}{\sigma_1\sigma_2}}
\d y \; 
\exp\left[-\frac{1}{4} \frac{\sigma_1\sigma_2}{\sigma_1\sigma_2-\xi(r_{12})} y^2\right]
.\label{eq:p12nu}
\end{equation}
This integral is of Laplace type 
$I=\int_1^{\infty} \d x \; e^{-\alpha \phi(x)} f(x)$, 
$\alpha=\frac{\delta_{\rm cr}^2}{\sigma_1\sigma_2}$ which for large $\alpha$
asymtotically accumulates at the lower integration boundary over the interval
$x \in [1,1+\Delta x], ~\Delta x = 2 \frac{\sigma_1 \sigma_2}{\delta_{\rm cr}^2}
\frac{\sigma_1\sigma_2+\xi(r_{12})}{(\sigma_1+\sigma_2)^2}$.
Asymtotic expansion is straightforward if one can expand $f(x)$ in the Taylor
series near $x=1$
\begin{equation}
f(x) \equiv 
\int_{\frac{\delta_{\rm cr}[2\sigma_2-x(\sigma_1+\sigma_2)]}{\sigma_1\sigma_2}}^{\frac{\delta_{\rm cr}[x(\sigma_1+\sigma_2)-2\sigma_1]}{\sigma_1\sigma_2}}
\d y \; e^{-\frac{1}{4} \frac{\sigma_1\sigma_2}{\sigma_1\sigma_2-\xi(r_{12})}
y^2}
\approx 2 \left(x-1\right)\frac{\delta_{\rm cr} (\sigma_1+\sigma_2)}{\sigma_1\sigma_2}
\exp\left[-\frac{1}{4} \frac{\delta_{\rm cr}^2}{\sigma_1\sigma_2}
\frac{(\sigma_1-\sigma_2)^2}{\sigma_1\sigma_2-\xi(r_{12})}\right],
\label{eq:fxTaylor}
\end{equation}
in which case we get
\begin{equation}
p(1,2) \approx \frac{2}{\pi} \frac{\sigma_1\sigma_2}{\delta_{\rm cr}^2}
\frac{[\sigma_1\sigma_2+\xi(r_{12})]^{3/2}}
{(\sigma_1+\sigma_2)^2 (\sigma_1\sigma_2-\xi(r_{12}))^{1/2}}
\exp\left[-\frac{1}{2} \delta_{\rm cr}^2  
\frac{\sigma_1^2+\sigma_2^2-2\xi(r_{12})}
{\sigma_1^2\sigma_2^2-\xi^2(r_{12})}\right].
\label{eq:p12gen}
\end{equation}

Two things are notable. First is the prefactor
$(\sigma_1\sigma_2)/\delta_{\rm cr}^2$. Second, we find
that for small correlations  the effect in the exponent
where small $\xi/\sigma_1\sigma_2$ is multiplied by
$\delta_{\rm cr}/\sigma_1\sigma_2$
dominates the correction from the prefactor. Thus, 
as a leading order approximation, we can account for small correlations
by factoring out the exponential correlation term 
from the original expression, with
the values of the field replaced by the threshold values.

In reality, the asymptotics in eq.\ (\ref{eq:p12gen}) do not give
an accurate approximation if correlations are strong $\xi(r_{12}) \to 
\sigma_1\sigma_2$, especially since our threshold parameter $\delta_{\rm cr}/\sigma$
may not be very large. This is definitely the case for a distribution of
identical objects at short distances, since then $\xi(r\to0)=\sigma^2$.
More accurately, the Taylor expansion of $f(x)$ in eq.(\ref{eq:fxTaylor}) 
is not suitable when the width of the relevant integration range
$\Delta y = \delta_{\rm cr} \Delta x (\sigma_1+\sigma_2)/(\sigma_1\sigma_2)$
exceeds the width of the Gaussian 
$\sqrt{\left[\sigma_1\sigma_2-\xi(r_{12}) \right]/(\sigma_1\sigma_2)}$.
In this case, however, the integration over $y$ can be extended to
$\pm\infty.$ With subsequent asymptotic analysis of the integral over $x$
this gives
\begin{equation}
p(1,2) \approx
\frac{1}{\sqrt{\pi}} \frac{\sqrt{\sigma_1\sigma_2}}{\delta_{\rm cr}}
\frac{\sqrt{\sigma_1\sigma_2+\xi(r_{12})}}{\sigma_1+\sigma_2}
\exp\left[-\frac{1}{4} \frac{\delta_{\rm cr}^2}{\sigma_1\sigma_2}  
\frac{(\sigma_1+\sigma_2)^2}{\sigma_1\sigma_2+\xi(r_{12})}
\right], ~~~~
\frac{2}{\sqrt{\sigma_1\sigma_2-\xi(r_{12})}}
\frac{\sigma_1\sigma_2+\xi(r_{12})}{\sigma_1+\sigma_2}
\gg \frac{\delta_{\rm cr}}{\sqrt{\sigma_1\sigma_2}} \gg 1
\label{eq:p12genhc}
\end{equation}

The general eqs.\ (\ref{eq:p12gen})-(\ref{eq:p12genhc}) 
are much simpler in the case when variances are
identical $\sigma_1^2=\sigma_2^2=\xi(0)$. Defining the cross correlation
coefficient $c(r_{12}) = \xi(r_{12})/\xi(0)$ and specifying accurately
the range of validity of eq.\ (\ref{eq:p12gen}) gives
\begin{eqnarray}
\EQN{eq:p12use1}
p(1,2) &\approx& \frac{1}{2 \pi} \nu^{-2} \; 
\Mvariable{A}\left[c(r_{12})\right]
\exp\left[-\frac{\nu^2}{1+c(r_{12})}\right] ~~~, ~~~~ \nu \gg 
\max\left(1,\frac{1+c(r_{12})}{\sqrt{1-c(r_{12})}} \right), \\
p(1,2) &\approx& \frac{1}{\sqrt{2 \pi}} \nu^{-1} \;
\Mvariable{B}\left[c(r_{12})\right]
\exp\left[-\frac{\nu^2}{1+c(r_{12})}\right] 
~, ~~~~ \frac{1+c(r_{12})}{\sqrt{1-c(r_{12})}} \gg \nu \gg 1 ~~,
\EQN{eq:p12use2}
\end{eqnarray}
where smooth functions $A(x)\equiv\sqrt{\frac{(1+x)^3}{1-x}}
\stackrel{x \to 0}{\longrightarrow}1$
and $B(x) \equiv \sqrt{\frac{1+x}{2}} \stackrel{x\to1}{\longrightarrow}1$.
As expected, $p(1,2)\stackrel{c \to 0}{\longrightarrow} p(1)p(2)$ and
$p(1,2)\stackrel{c\to1}{\longrightarrow} p(1)$.
It is important to note that the probability is additionally
enhanced by $\nu=\delta_{\rm cr}/\sigma$ when correlations are strong.

Finally, we combine \Eqs{eq:p12use1}{eq:p12use2}
into the uniform approximation
\begin{equation}
p(1,2) \approx \frac{1}{2 \pi} \nu^{-2} \;
\Mvariable{C}\left[c(r_{12}),\nu \right]
\exp\left[-\frac{\nu^2}{1+c(r_{12})}\right]
~, ~~~~ \nu \gg 1,
\label{eq:p12uni}
\end{equation}
with the help of an interpolating function, 
$\Mvariable{C}(x,\nu)$, such that 
$\Mvariable{C}(0,\nu)=1,~\Mvariable{C}(1,\nu)=\nu\sqrt{2\pi}$.
The choice
\begin{equation}
\Mvariable{C}(x,\nu) = \frac{\nu\sqrt{\pi}\sqrt{(1+x)^3}}
{\left(\nu\sqrt{\pi}-1\right) \sqrt{1-x}+(1+x)}
\EQN{Cipol}
\end{equation}
reflects both the details of the functions $A(x)$ and $B(x)$ and of the 
transition between \Eqs{eq:p12use1}{eq:p12use2}.

In  the weak correlation regime, the formula (\ref{eq:p12uni}) coincides with
the classic result of Kaiser (1984). 
At the same time in the strong correlation
regime, the result (\ref{eq:p12uni}) shows that the correlation between regions
of high density is additionally enhanced by $\sqrt{2 \pi} \nu$ factor.
Although our result is rigorous for the points of high excursions of the field at all separations $r$, the interpretation of the last
regime in terms of peak, or object, correlation,
is questionable at $r < R_1+R_2$ when the two high-density points  
likely belong to the same peak. 

\end{document}